\documentclass[useAMS,usenatbib,usegraphicx]{mn2e}
\usepackage{aas_macros}
\usepackage[dvips]{graphicx}
\usepackage{longtable}
\usepackage{color} 
\usepackage{colortbl}   
\usepackage{lscape}
\usepackage{multirow}
\usepackage[figuresright]{rotating}
\usepackage{bigstrut}
\usepackage{amssymb}
\newcommand{\Ha}{H$\alpha$ }

\newcommand{\naco}{NaCo }
\definecolor{LightGray}{gray}{0.8}

\title[The innermost GCs of M87]{The innermost globular clusters of M87\thanks{Based on European Southern Observatory (ESO) Very Large Telescope (VLT) programs 074.B-0404(A) and 076.B-0493(A) and observations made with the NASA/ESA \emph{Hubble Space Telescope}, obtained from the Data Archive at the Space Telescope Science Institute, which is operated by AURA, Inc., under NASA contract NAS 5-26555 }}

\author[M. Montes et al. ]{M. Montes$^{1,2}$\thanks{E-mail: mireia.montes.quiles@gmail.com (MM)}, J. A. Acosta-Pulido$^{1,2}$, M. A. Prieto$^{1,2}$, J. A. Fern{\'a}ndez-Ontiveros$^{3}$ \\ 
$^{1}$ Instituto de Astrof{\'i}sica de Canarias (IAC), V{\'i}a L{\'a}ctea s/n, La Laguna, E-38200, Spain\\
$^{2}$ Departamento de Astrof{\'i}sica, Facultad de F{\'i}sica, Universidad de La Laguna, Astrof{\'i}sico Fco. S{\'a}nchez s/n, La Laguna, E-38207, Spain \\
$^{3}$ Max-Planck-Institut f\"ur Radioastronomie (MPIfR), Bonn, D-53121, Germany}

\begin{document}

\date{Accepted 2014 May 9. Received 2014 May 9; in original form 2013 May 31}
\pagerange{\pageref{firstpage}--\pageref{lastpage}} \pubyear{2014}
\maketitle
\label{firstpage}

\begin{abstract}
We present a comprehensive multiwavelength photometric analysis of the innermost ($\sim3\times3$ kpc$^2$) $110$ globular clusters (GCs) of M87. Their spectral energy distributions (SEDs) were built taking advantage of new ground-based high resolution near-IR imaging aided by adaptive optics at the {Very Large Telescope} (VLT) combined with \emph{Hubble Space Telescope} (\emph{HST}) ultraviolet--optical archival data. These GC SEDs are among the best photometrically sampled extragalactic GC SEDs. To compare with our SEDs we constructed equally sampled SEDs of Milky Way GCs. Using both these Milky Way cluster templates and different stellar population models, ages of $>10$ Gyr and metallicities of [Fe/H] $\sim -0.6$ dex are consistently inferred for the inner GCs of M87. In addition, the metallicity of these GCs is low ($\Delta$[Fe/H] $\sim 0.8$ dex) compared to that of their host galaxy. These results agree with the idea that the GC formation in M87 ceased earlier than that of the bulk of the stars of the central part of the galaxy. The ages of the inner GCs of M87 support the idea that these central parts of the galaxy formed first. Our data do not support evidence of recent wet merging.

\end{abstract}

\section{Introduction}\label{Intro}

Globular clusters (GCs) are among the oldest stellar components in galaxies. Their stellar populations have been widely recognized as holding important clues about the formation of both GCs and their host galaxy. It is believed that GCs formed in episodes of intense star formation, i.e.\ the major star-forming episodes that shaped galaxies \citep[see][and references therein]{Brodie2006}. Therefore, the properties of both systems, especially their metallicities, are correlated \citep[e.g.][]{Brodie2006}. 

Our present understanding of massive galaxy formation suggests a two-step scenario: the central part is the relic of the early stages of formation, and their extended halo results from minor merging \citep[e.g.][]{Trujillo2011}. In this sense, GC systems may provide constraints for this scenario. For example, the presence of GC formation may help to distinguish between two types of minor mergers in a galaxy: dry (little or no gas) or wet (with gas). For instance, the recent discovery of young star clusters in the central regions of elliptical galaxies \citep[e.g. NGC~1052, ][]{Fernandez-Ontiveros2011} is related to the gas provided by past merger events \citep[e.g. ][]{Kaviraj2010, Kaviraj2010a}. 

In order to obtain accurate estimates of age and metallicity it is crucial to have access to a large wavelength range (ideally including UV, optical and NIR). This is valid even for old ages if other parameters (e.g.\ metallicity) can be fixed a priori \citep{Anders2004}. Examples of this can be found in a series of papers by \citet[][and references therein]{Wang2010}, where the  M$31$ GCs were derived from model fitting of the spectral energy distributions (SEDs). The NIR is mainly dominated by stars populating the red giant branch, hence by using the optical/NIR combination we have a better chance to break the age--metallicity degeneracy inherent in optical wavelengths \citep[e.g.][]{Kissler-Patig2002}.

In this paper, we focus on M87, the giant elliptical galaxy near the core of the Virgo cluster, as a representative example of a massive galaxy. The GC system of M87 has been one of the most carefully studied extragalactic systems. Its numerous GCs \citep[$\sim14000$,][]{Tamura2006} and the proximity of the galaxy \citep[D=$16.1$ Mpc,][]{Blakeslee2001} make it a perfect target for understanding the formation of massive galaxies. 
The colour distribution of M87 GCs reveals two distinct subpopulations in optical colours \citep[e.g.][]{Peng2009, Waters2009} defined as \emph{blue} and \emph{red}. The red clusters are more centrally concentrated than the blue clusters \citep[e.g. ][]{Kundu1999, Harris2009,Strader2011}. Furthermore, the radial distribution of the red GCs closely matches that of the galaxy light \citep{Forte2012}. However, the difference in colours does not translate into a difference in age \citep{Cohen1998, Jordan2002}, but in a difference in metallicity. \citet{Kundu1999} using \emph{HST} imaging found that the two peaks in $V-I$ corresponded to [Fe/H] $=-1.41$ dex and [Fe/H] $=-0.23$ dex. This difference in metallicity was confirmed by other studies \citep[e.g.][]{Jordan2002, Kaviraj2007}. \citet{Cohen1998} acquired and analysed spectra of $150$ GCs in M87 and estimated the metallicity with spectroscopic indices. They found that the M87 GCs show a large range in metallicity, from metal-poor to super-solar. The metallicity distribution of their sample is marginally bimodal in metallicity and has an extended metal-rich tail. Their mean [Fe/H] is $-0.95$ dex.

Some studies use the capabilities of the \emph{Hubble Space Telescope} (\emph{HST}) to obtain high spatial resolution data of the clusters in other wavelength ranges. \citet{Sohn2006} used STIS imaging of four fields in the inner regions of M87 to obtain near and far-ultraviolet (FUV) photometry finding enhanced UV fluxes in M$87$ GCs. This UV excess is thought to be caused by the presence of the so-called extreme horizontal branch (EHB), populated by low-mass, helium-burning stars \citep[see][for a review]{O'Connell1999}. \citet{Kaviraj2007} compiled photometry of M87 GCs spanning the widest wavelength range to date, from the UV to the $I$-band. Their derived ages were significantly in excess of the currently accepted age of the Universe due to this UV excess. Separately, there have been studies in the near-infrared (NIR) range investigating the properties of this GC system \citep[e.g.][]{Kundu2007,ChiesSantos2011b, ChiesSantos2011a} 

 \begin{figure*}
  \begin{center}
\renewcommand{\multirowsetup}{\centering}
    \begin{tabular}{@{}c@{}c@{}c@{}}
    \includegraphics[height=0.35\textheight]{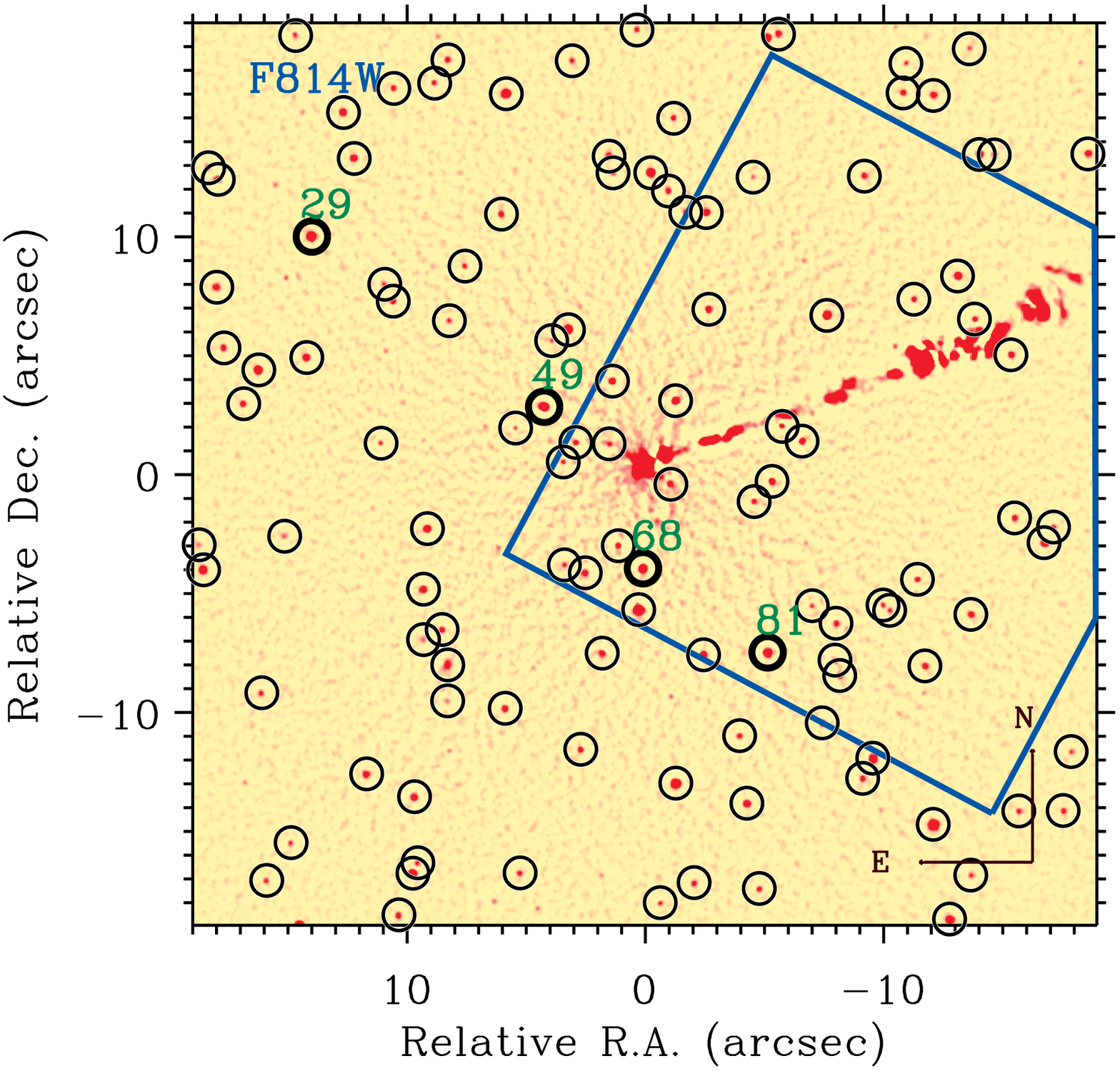} &
    \includegraphics[height=0.35\textheight]{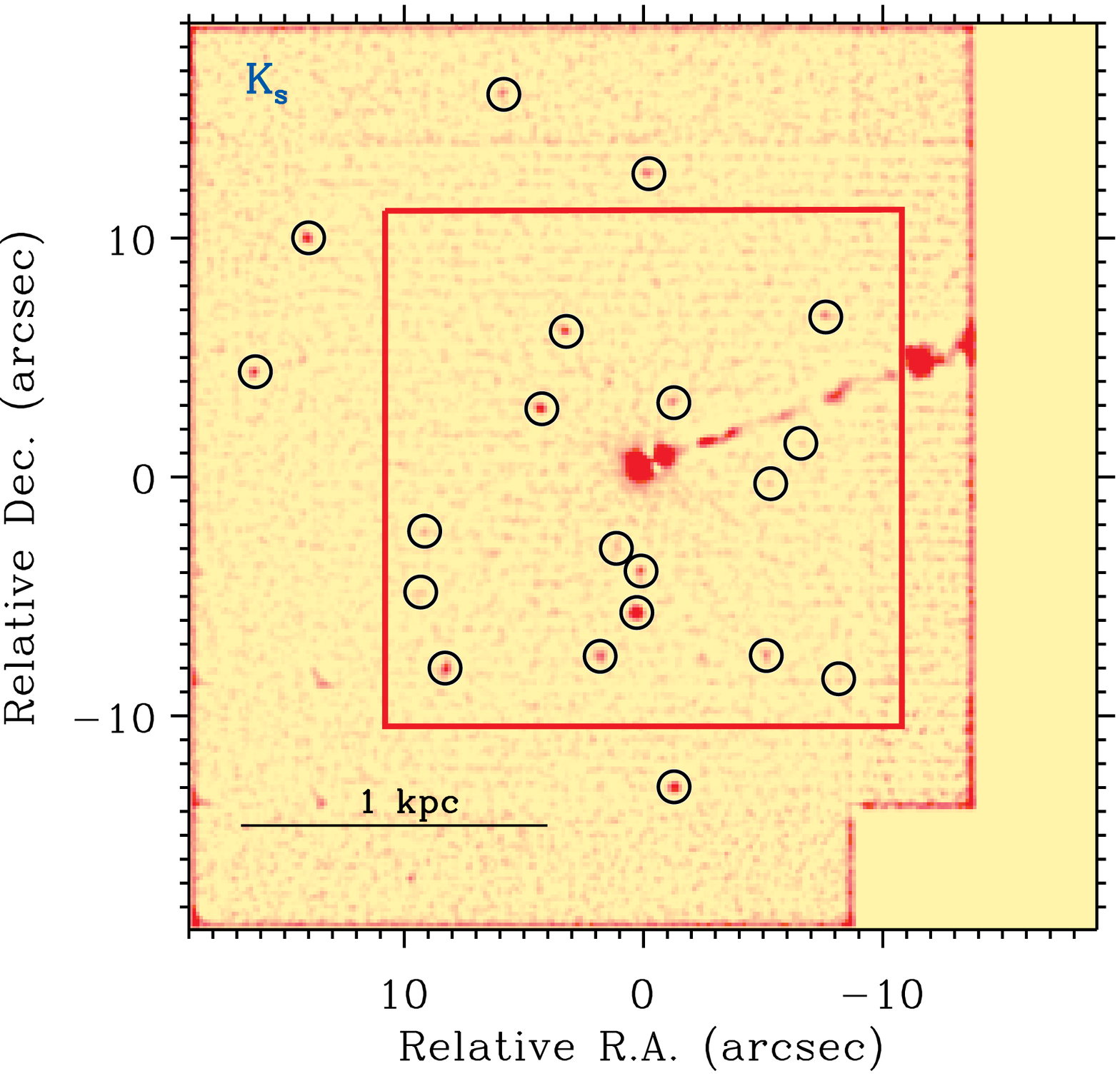} \\ 
    \includegraphics[height=0.35\textheight]{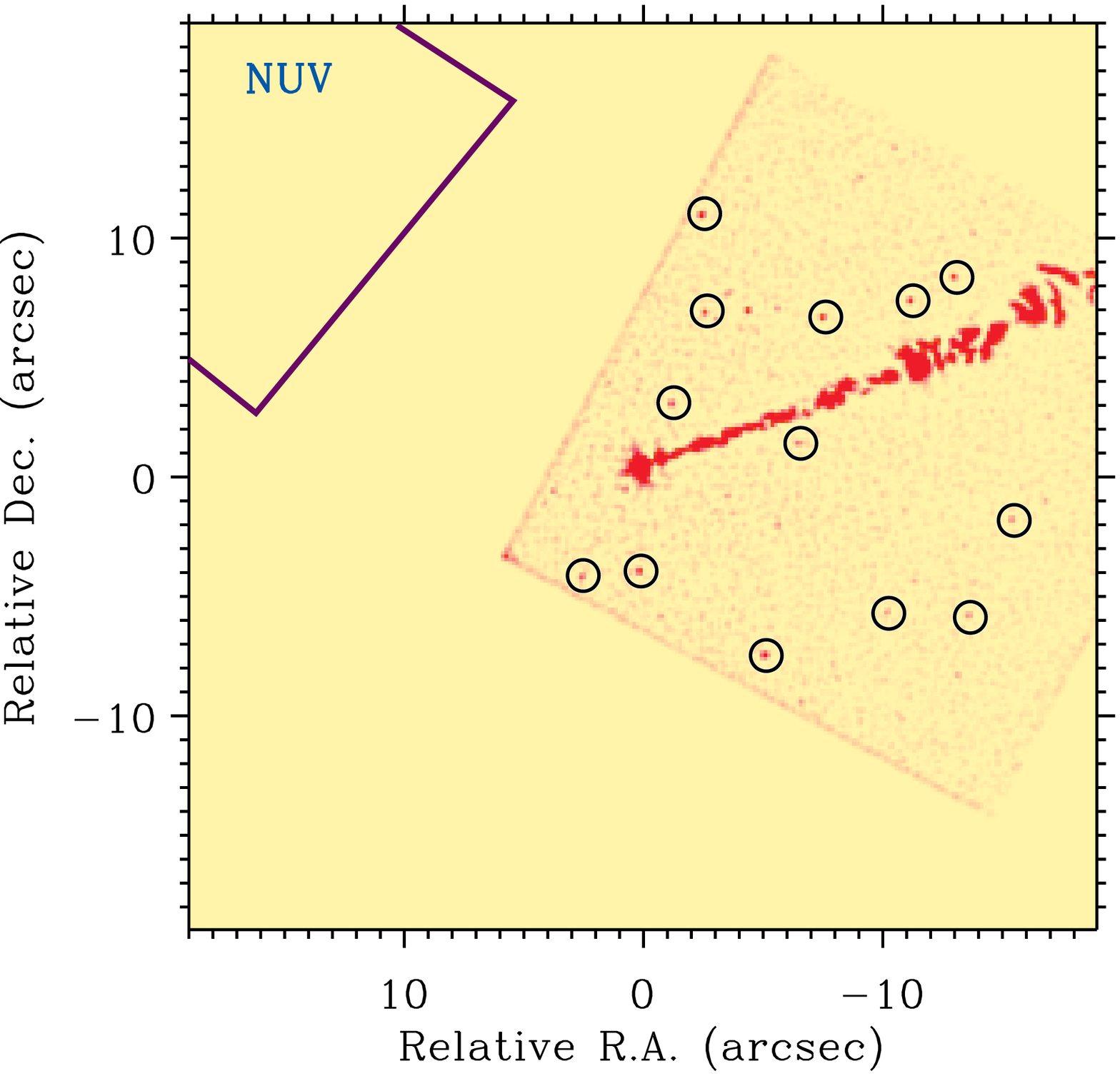} &
    \includegraphics[height=0.35\textheight]{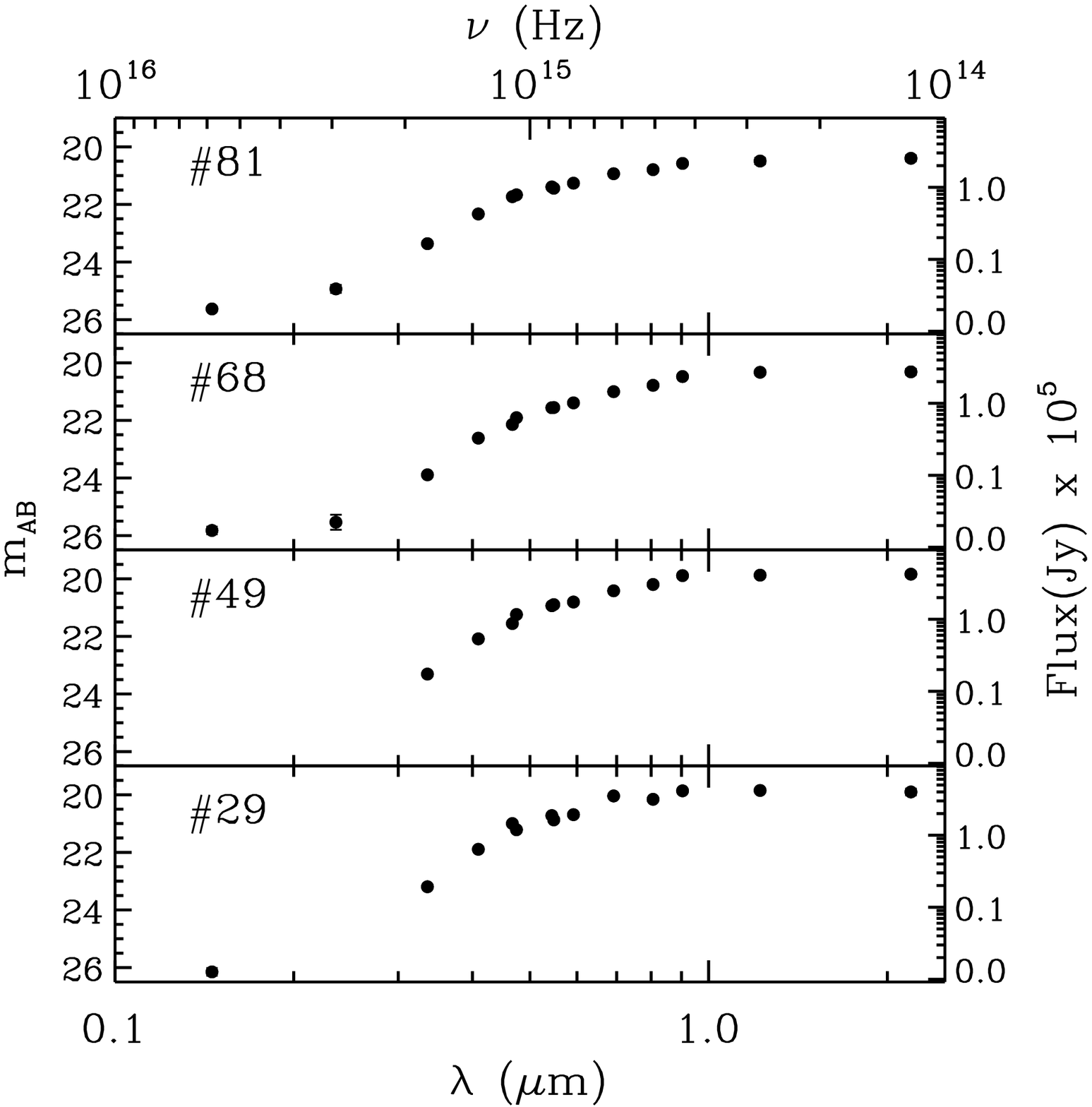} \\
    \end{tabular} 
   \caption{The upper left panel shows the unsharp-masked F$814$W \emph{HST} image smoothed with a median filter of $0\farcs25$ for the central $~38\times38$ $arcsec^2$ of M$87$ corresponding to the $J$-band field of view. The detected sources are marked by circles. The upper right and bottom left panels show the corresponding unsharp-masked $K_s$ and NUV STIS images respectively with the detected sources marked. The blue square in F$814$W corresponds to the FOV of the NUV image. The red square in $K_s$ corresponds to the FOV of the $K_s$ image of $2005$ and the purple square in NUV corresponds to the F25SFR2 image of $2002$. Bottom right panel: SEDs of a subsample of $4$ GCs, marked and numbered in the F$814$W image.}\label{iplussed}
  \end{center}
 \end{figure*}

This paper presents the first multiband analysis of the GCs in the inner regions of M87 that also includes the NIR range. 
Advances in IR imaging, improved with adaptive optics, allow us to measure the GCs in the central parsecs of galaxies despite the intense background light of the galaxy. In this paper, we will explore the ages and metallicities of these GCs. This may provide a valuable tool for understanding the early phases of the formation and the consequent evolution of both GCs and their host galaxy.

The structure of this paper is as follows. In Section \ref{data}, we present the multiwavelength data used here. Section \ref{photometry} gives details on the photometry and the SEDs. We explore the age and metallicity distributions of the GCs of M87 in Section \ref{agesandmets} with an improved accuracy due to the inclusion of NIR quality imaging. In Section \ref{discussion}, we compare our results with previous studies and interpret them in the context of galaxy formation.  

\section{Data}\label{data}

 \begin{table*}
  \centering
   \begin{tabular}{@{}lllcrccc@{}}\hline
Filter & $\lambda$ (\AA{})& $\Delta \lambda\,(\AA)$ & Instrument & Pixel scale ($\arcsec/pixel$) & Exp. Time ($s$) & Date & Programme      \\ 
\hline
  F25SRF2    &            &              &                     &                 &           &               & \\
   FUV-MAMA  &  $1456.6$ & $120.8$       & STIS@\emph{HST}     &  $0.025$       &  $3526$   & $17/05/1999$  & $8140$ \\
  F25SRF2    &            &              &                     &                 &           &               & \\
   FUV-MAMA  &  $1456.6$ & $120.8$       & STIS@\emph{HST}     &  $0.025$       &  $2680$   & $14/02/2002$  & $8643$ \\ 
  F25QTZ     &            &              &                     &                 &           &               & \\
   NUV-MAMA  &  $2355.3$ & $419.8$       & STIS@\emph{HST}     &  $0.025$       &  $2372$   & $17/05/1999$  & $8140$ \\ 
  F336W      &  $3359.5$ & $204.5$       & WFPC2@\emph{HST}    &  $0.046$/$0.1$ &  $28800$  & $25/12/2000$  & $8587$  \\ 
  F410M      &  $4092.7$ & $93.8$        & WFPC2@\emph{HST}    &  $0.046$/$0.1$ &  $19200$  & $07/02/2001$  & $8587$  \\ 
  F467M      &  $4670.2$ & $75.3$        & WFPC2@\emph{HST}    &  $0.046$/$0.1$ &  $7200$   & $28/12/2000$  & $8587$ \\ 
  F475W      &  $4745.3$ & $420.1$       & ACS/WFC@\emph{HST}  &  $0.05$        &  $750$    & $19/01/2003$  & $9401$ \\ 
  F547M      &  $5483.9$ & $205.5$       & WFPC2@\emph{HST}    &  $0.046$/$0.1$ &  $7200$   & $10/02/2001$  & $8587$ \\
  F555W      &  $5442.9$ & $522.2$       & WFPC2@\emph{HST}    &  $0.046$/$0.1$ &  $2430$   & $03/02/1995$  & $5477$ \\
  F606W      &  $5919.4$ & $672.3$       & ACS/WFC@\emph{HST}  &  $0.05$        &  $28500$  & $24/12/2005$  & $10543$   \\ 
  F658N      &  $6590.8$ & $29.4$        & WFPC2@\emph{HST}    &  $0.046$/$0.1$ &  $13900$  & $09/05/1996$  & $6296$ \\ 
  F702W      &  $6917.1$ & $586.7$       & WFPC2@\emph{HST}    &  $0.046$/$0.1$ &  $280$    & $23/01/1995$  & $5476$ \\
  F814W      &  $8059.9$ & $653.0$       & ACS/WFC@\emph{HST}  &  $0.05$        &  $72000$  & $24/15/2005$  & $10543$  \\ 
  F850LP     &  $9036.4$ & $527.2$       & ACS/WFC@\emph{HST}  &  $0.05$        &  $1120$   & $19/01/2003$  & $9401$ \\ 
  $J$        &  $12650.0$ & $2500.0$     & NACO@VLT            &  $0.027$       &  $784$ & $23/01/2006$  & $076.B-0493(A)$ \\
  $K_s$      &  $21800.0$ & $3500.0$     & NACO@VLT            &  $0.027$       &  $300$ & $23/01/2006$  & $076.B-0493(A)$ \\
  $K_s$      &  $21800.0$ & $3500.0$     & NACO@VLT            &  $0.027$       &  $680$ & $20/01/2005$  & $074.B-0404(A)$ \\ \hline

   \end{tabular} 
   \caption{Description of the set of images used in this paper.}\label{datadetails}
 \end{table*}

\subsection{NIR Data}
The NIR data used here consist of high spatial resolution images in the $J$ and $K_s$ bands acquired with the Very Large Telescope (VLT) using the Nasmyth Adaptive Optics System plus the Near-Infrared Imager and Spectrograh (NAOS + CONICA, NaCo). The inner $\sim38\times 38$ arcsec$^2$ ($\sim3\times3$ kpc$^2$, $1$ arcsec$=78$ pc at the distance of M87, \citealt{Blakeslee2001}) of M$87$ were observed. Two images in the $K_s$ band were taken in $2005$ and $2006$ targeting a field of view of $\sim22 \times22$ and $\sim33\times 38$ arcsec$^2$, respectively. The achieved resolution was $0\farcs27$ ($J$) and $0\farcs12$ and $0\farcs19$ ($K_s\;2005$ and $K_s\;2006$) measured as the full width at half maximum (\textsc{FWHM}) of the most compact source in each image. The \naco data reduction was performed using the \textsc{eclipse} package provided by ESO \citep{Devillard1997}. Photometric calibration of the images made use of standard stars taken along with the science frames. 
More details of the data can be found in Table \ref{datadetails}. We used the $K_s\,2005$ image for the central part (marked as a red square in the left upper panel of Figure \ref{iplussed}), and $K_s\,2006$ for the outer parts.

\subsection{Optical and UV Data}
In order to explore the inner regions of M87 with data of comparable resolution, the NIR data were supplemented with \emph{HST} archival data covering the UV--optical range. The optical images are from the WFPC2 and the ACS.

The images from the WFPC2 were taken to observe the jet and the active nucleus with the PC camera, so most of our sources fall within this chip. STIS FUV MAMA (F25SFR2) and NUV MAMA (F25QTZ) images of the central region of M87 were also taken from the archive although they only overlap with half of the \naco $J$-band field of view (see Figure \ref{iplussed}). We also added the F25SFR2 for an adjacent field that covers partially our field of view.
The \emph{HST} images were combined using \textsc{multidrizzle} \citep{Fruchter2002} to achieve the nominal resolution of each camera. \textsc{multidrizzle} combines the aligned individual exposures, rejects outliers and removes the geometric distortions. 
More details about the dataset are listed in Table \ref{datadetails}.

\section{Photometry}\label{photometry}

Image alignment of the whole data set is based on the identification of three sources present in all the filters.
We used the deepest image, ACS/F$814$W, matching the field of view of the NaCo $J$ image, for source detection. To enhance the contrast, the unsharp-masking method \citep[][]{Sofue1993} was used, as illustrated in Figure \ref{iplussed}. A median box filter of size $\sim2\times$FWHM of the sources was applied to all images and then the filtered images were subtracted from the original ones to remove the diffuse galaxy light. The lists of detected sources in each image were cross-correlated to create the final catalogue. Table \ref{tablaradec} lists the right ascension and declination of the sources relative to the nucleus of M87, and their correspondence with \citet{Kundu1999} optical and \citet{Sohn2006} UV data.
Detection and photometry of the sources were carried out using \textsc{SExtractor} \citep{Bertin1996} on the original images. The detection algorithm recovered a total of $129$ sources in the F$814$W image using a threshold of five connected pixels at a $3\sigma$ significance level. 
The photometry was carried out by means of aperture photometry with \textsc{SExtractor}. In our case, an aperture of the size $\sim$\textsc{FWHM} of the sources was used ($\sim0\farcs16$ in the optical, $\sim0\farcs13$ in the UV and $\sim0\farcs25$ in the IR). 

To determine the completeness and to correct the contamination produced by the strong background light of the galaxy, we run a number of simulations locating artificial GCs on the images. This correction, or bias, is a position-dependent correction that accounts for the galaxy light, as well as for aperture corrections. This is obtained in the completeness tests. Further details on these tests are provided in Appendix \ref{simulations}. For illustration, Table \ref{tablaupp} shows the upper limits in magnitude defined as $68$ per cent of completeness for three radial distances from the nucleus of M$87$.\\

\begin{table*}  
 \begin{center}
  \begin{tabular}{ccccccccc}\hline
     Distance &    FUV &     NUV &     F336W &    F410M &  F467M &   F475W &   F555W &  \\
      kpc &  \multicolumn{8}{c}{\cellcolor[gray]{0.8}[mag AB]} \\ \hline
      $ 0.5  $&$ 26.33 $&$ 26.12 $&$ 24.63 $&$ 23.45 $&$ 23.13 $&$ 24.05$& $ 24.01$& \\ 
      $  1  $&$ 26.37 $&$ 26.21 $&$ 24.92 $&$ 23.61 $&$ 23.19 $&$ 24.15$& $ 24.28$& \\
      $1.5  $&$ 26.39 $&$ 26.87 $&$ 25.08 $&$ 24.36 $&$ 23.47 $&$ 24.20$& $ 24.38$& \\ \hline
           & F547M   &  F606W &  F658N   & F702W  & F814W    & F850LP    & $J$	& $K_s$ \\
           &  \multicolumn{8}{c}{\cellcolor[gray]{0.8}[mag AB]}\\ \hline
      $ 0.5 $ &$ 24.01 $&$24.35 $&$ 21.92 $&$ 22.97 $&$ 23.48 $&$ 21.95 $&$ 21.38$& $ 20.76 $ \\
      $  1  $ &$ 24.29 $&$24.78 $&$ 21.93 $&$ 23.17 $&$ 23.60 $&$ 22.24 $&$ 21.40$& $ 21.04 $ \\
      $1.5  $ &$ 24.53 $&$24.92 $&$ 21.94 $&$ 23.53 $&$ 23.80 $&$ 22.29 $&$ 21.44$& $ 21.04 $ \\ \hline
   \end{tabular} \caption{$68$ per cent completeness magnitudes for the $15$ filters for three radial distances to the centre of M87. $m-M=31.03$}\label{tablaupp} 
 \end{center}
\end{table*}

From the initially detected $129$ sources in the \emph{HST}/F$814$W image, only those located in regions with completeness (defined in Appendix \ref{simulations}) higher than $68$ per cent ($1\sigma$) in the simulations were kept. This leaves us with a total of $115$ sources. For each of these sources, we built a spectral energy distribution from the UV to the NIR. Photometric values in the SEDs were measured in the images after correcting for foreground extinction, $E(B-V)=0.022$ \citep{Schlegel1998}, using the \citet{Cardelli1989} extinction law. Four examples of these SEDs are shown in the bottom right panel of Figure \ref{iplussed}. 
The data for the $115$ sources are presented in Table \ref{tablacums} containing the photometry of the $15$ filters from the FUV to the NIR in AB magnitudes, together with the errors. There are five objects that are probably not GCs (as explained in Appendix \ref{datam87}) and have also been removed from the analysis. The final sample contains $110$ objects. The photometry listed in Table \ref{tablacums} includes the additional correction by the bias derived in our simulations while Table \ref{tablacumsnobias} does not include this correction. 

 \begin{figure}
  \begin{center}
    \includegraphics[height=0.35\textheight]{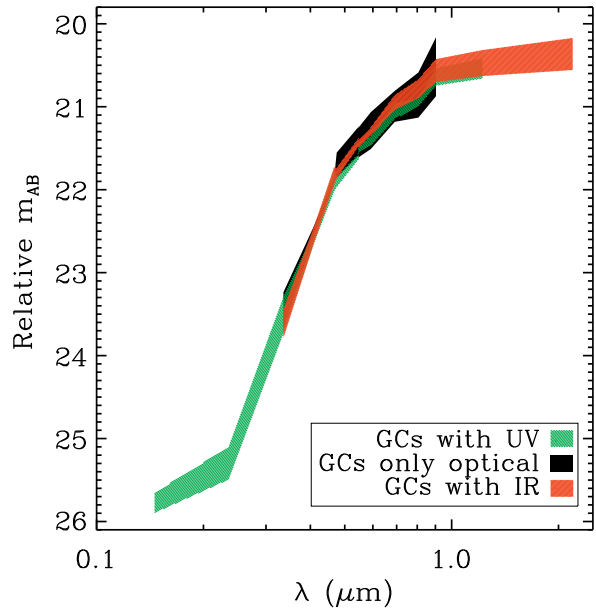} 
    \caption{Average SEDs for the three groups defined within the GCs of M87 with those emitting in the UV (green), in the IR (orange) and in the optical (black).}\label{promedios}
  \end{center}
 \end{figure}
 
\subsection{Spectral Energy Distributions of M87 GCs}\label{SED}

Not all our GCs are fully covered photometrically from the UV to the IR. This raises some questions about whether our inferences can be generalized in those cases where we do not have a complete photometric coverage. For this reason, we have built three representative average SEDs of M87 GCs. The averages were defined as three separate groups: {\it i)} clusters with only optical data ($16$ clusters), {\it ii)} clusters with IR and optical data ($17$ sources, but eight with $K_s$), and {\it iii)} clusters with UV and optical data ($11$ sources, only three with NUV and FUV). For each group, an average SED is created by normalizing individual SEDs using the total luminosity integrated upon the common optical bands. In order to avoid biases due to the different exposure times of the bands, we decided to exclude the clusters that do not have information in all of the filters considered.

 \begin{figure*}
  \begin{center}
    \includegraphics[height=0.4\textheight]{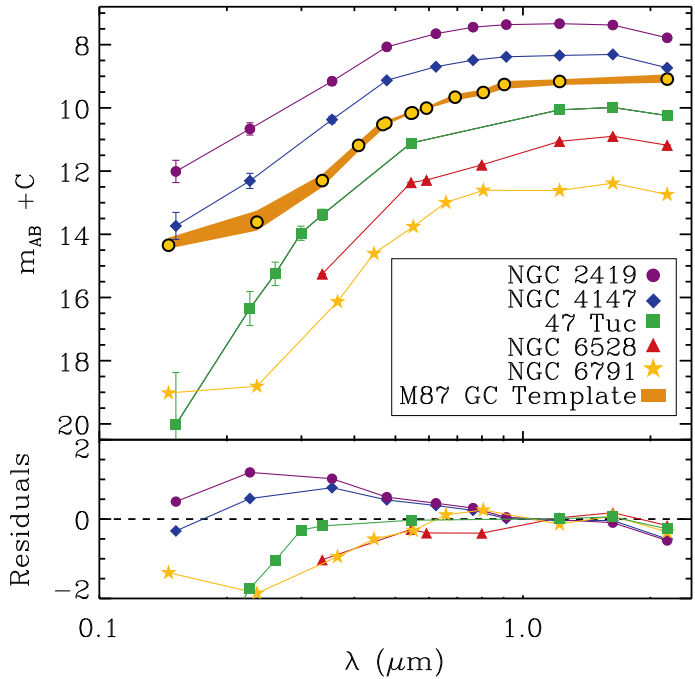} 
    \caption{Upper panel: Comparison of the average SED for M87 GCs with some MW clusters. The metallicities ([Fe/H]) for the MW GCs are: $-2.14$ for NGC~2419 (purple), $-1.80$ for NGC~4147 (blue), $-0.70$ for 47 Tuc (green), $-0.10$ for NGC~6528 (red) and $+0.45$ for NGC~6791 (yellow). The solid orange polygon corresponds to the average template SED for M87 GCs. Bottom panel: Residuals of the difference between the average of M87 GCs and the MW clusters. The mean metallicity for M87 GCs is $Z\sim0.004$ ([Fe/H]$\sim -0.7$, derived in Section \ref{VIcolour})}\label{figgalgc}
  \end{center}
 \end{figure*}
 
In Figure \ref{promedios}, we show the three average distributions. The shaded regions represent values between the first and third quartiles. The narrowness of these regions illustrates the similarity of the SEDs of the clusters independently of their photometric coverage. Furthermore, the averaged SEDs show a similar shape, suggesting that the mean properties of the clusters do not differ noticeably. Therefore, we created a unique template SED, ranging from the UV to the NIR, for the clusters of M87 combining the distributions of Figure \ref{promedios}. The final template SED for M87 GCs is shown in Figure \ref{figgalgc}.

\section{Age and metallicity determination}\label{agesandmets}

\subsection{Comparison with Galactic clusters: a qualitative approach to age and metallicity}\label{compGGCs}

Taking advantage of the wide wavelength range covered in this work, we compared the average SED with the SEDs of a number of  Galactic clusters. Deriving SEDs for objects whose properties can be measured by other methodologies provides a valuable tool for comparison. The chosen clusters are: $47$ Tuc, NGC~6528, NGC~4147 and NGC~2419. These Galactic GCs (GGCs) span a wide range of metallicities and inhabit different Galactic environments. The ages and metallicities of these clusters are listed in Table \ref{tablafitmw}. For further details on the photometry and properties of these GGCs see Appendix \ref{sedGC}. 

As shown in Figure \ref{figgalgc}, $47$ Tuc is the cluster that best reproduces the average SED of M87 GCs, except in the UV. The fact that the average M87 GC SED replicates the optical--NIR of $47$ Tuc indicates that the mean metallicity of the clusters of M87 is similar to that of this GGC, i.e.\ [Fe/H] $\sim-0.7$. However, the difference in the UV bands can be explained by the presence of a UV excess ascribed to a significant fraction of EHB stars in M87 GCs \citep{Sohn2006}, which is unseen in $47$ Tuc.

Furthermore, we have compared with the open cluster NGC~6791. This old ($8.3$ Gyr) massive open cluster with supersolar metallicity ([Fe/H]$=0.45$) has hot sources identified as EHB stars that make it a good comparison regarding the UV-enhanced properties of M87 GCs \citep[see][and references therein]{Buzzoni2012}. The photometry was taken from table 1 of \citet{Buzzoni2012}. The observed disagreement in the UV fluxes between M87 GCs and NGC~6791 (although both clusters possess EHB stars) could be explained by a variety of scenarios, including age difference, a larger relative number of HB stars in M87 GCs, or a lower metal abundance in M87 HB stars. 

\subsection{Empirical metallicity determination based on the $V-I$ colour}\label{VIcolour}
 
\begin{figure}
  \begin{center}
    \includegraphics[height=0.35\textheight]{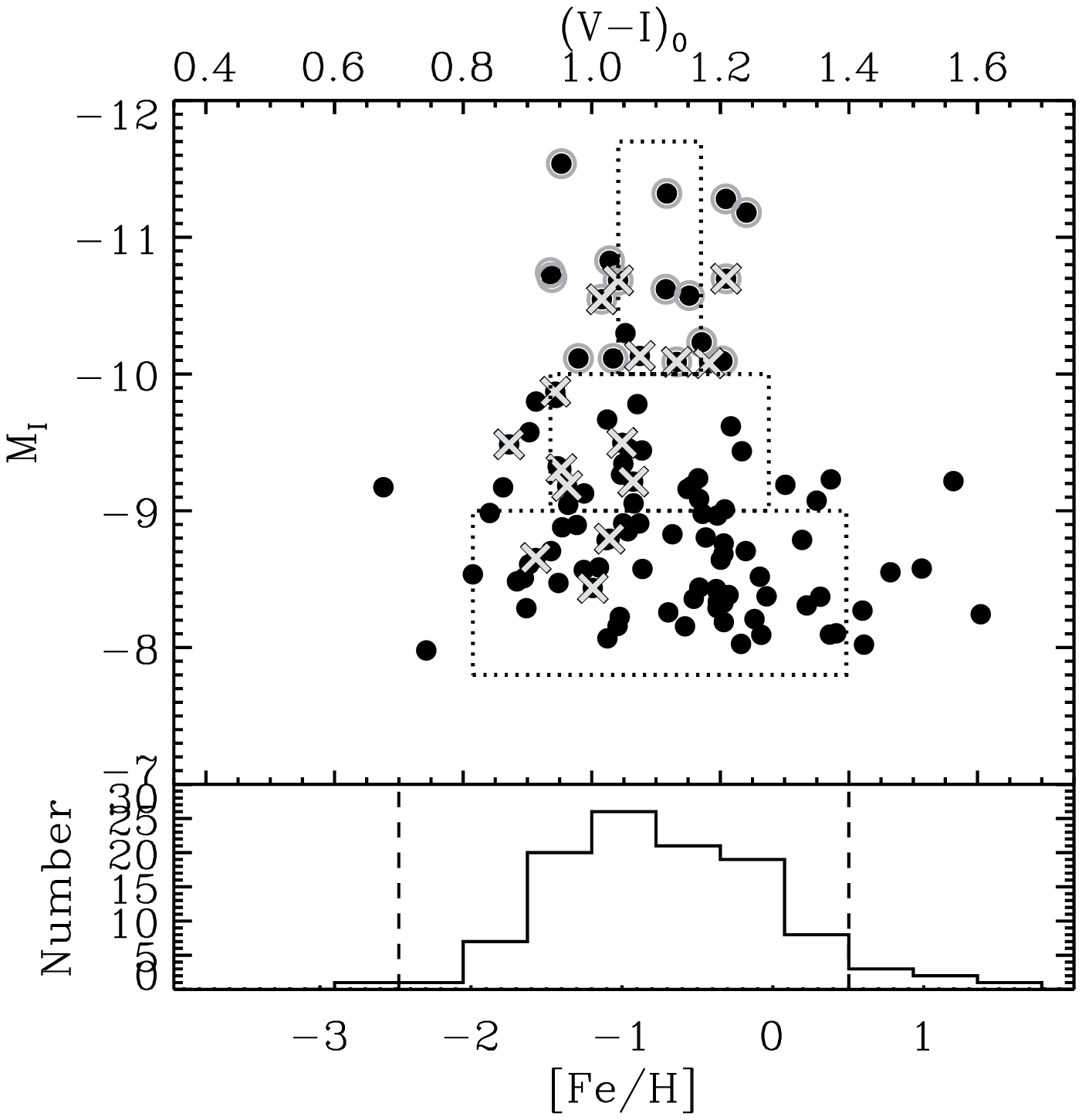} 
    \caption{Upper panel: colour--magnitude diagram of the $110$ M87 GCs. Lower panel: histogram of the $V-I$ colour of the clusters. The open grey circles are the clusters with IR photometry and the crosses mark the clusters with UV counterparts. The metallicity values indicated in the bottom $x$-axis were estimated using the linear transformation from $V-I$ to metallicity provided by \citet{Sohn2006}. The dotted boxes represent magnitude intervals in $y$ and the error of the median in $x$.}\label{fighistocol}
  \end{center}
 \end{figure}

A method of estimating the metallicity of a stellar population, assuming a certain age homogeneity as in the case of M87 GCs \citep{Jordan2002}, is to calculate its optical colours. We derived the $V-I$ (Vega system) colour of the GCs in the following way. The transformation from the F$606$W and F$814$W filter ST magnitudes to Vega magnitudes was taken from Table $10$ of \citet{Sirianni2005} and then to the standard $V$ and $I$ filters using equation 3 from \citet{DeGraaff2007}. Figure \ref{fighistocol} shows the colour--magnitude diagram and the histogram of the $V-I$ colour for our $110$ GCs. Different symbols are used to represent clusters belonging to the three groups defined in Section \ref{SED}, depending on the photometric coverage. The mean colour of our sample is $V-I=1.1$ mag in agreement with the mean colour in \citet{Kundu1999}, \citet{Kundu2007} and \citet{Peng2009}. To test if this colour distribution can be described by a single-peaked distribution, we calculated the probability that this hypothesis is true using a chi-squared distribution. The single-peaked hypothesis cannot be rejected with a confidence larger than $80\%$. In Fig. \ref{fighistocol},  we also draw the median errors in different bins of magnitudes to show that the distribution in $V-I$ and its scatter is compatible with a single colour (metallicity).

We also included the colour--metallicity transformation derived by \citet{Sohn2006} (their appendix B). They derived a linear relationship between the $V-I$ colour and [Fe/H], including information from Galactic GCs and two elliptical galaxies (NGC$1399$ and M$87$). The dashed vertical lines specify the interval where this relationship is valid. Contrary to other studies, we find no obvious bimodality in our data. A likely explanation is that our study is restricted to the innermost regions of M87, where the red subpopulation is more concentrated \citep{Strader2011} and the blue population is not prominent enough. The median of our distribution is $V-I=1.1$ mag ([Fe/H] = $-0.75\pm0.70$), compatible with the mean colour of the GCs in \citet{Kundu1999} ($V-I=1.09$). This metallicity agrees with the results obtained from the qualitative comparison with GGCs in Section \ref{compGGCs}.

\citet{Peng2006} found that a non-linear relationship better describes the colour--metallicity relationship of globular clusters. Using their non-linear relationship (for $42$ clusters with F$475$W-F$850$LP colours), we derived a median metallicity of [Fe/H] = $-0.61\pm0.40$, which is compatible with our previous estimate to within 2$\sigma$.  

\subsection{Absence of recent star formation}\label{Halpha}

Motivated by the discovery of very young stellar clusters in the core region of the elliptical galaxy NGC~1052 \citep{Fernandez-Ontiveros2011}, and also taking into account the UV enhancement of the GCs in M87, we searched for \Ha emission associated with these sources. Using a WFPC2 F$658$N image we measured the \Ha equivalent width (EW) of our clusters. The continuum level was inferred from a linear interpolation between F$606$W and F$702$W in magnitude. In those cases where the F$702$W photometry was not available we used only the F$606$W flux. All the globulars show an EW in \Ha compatible with zero within errors. In fact, although there are pieces of evidence for a merger event towards the centre of M87 \citep[e.g.][]{Romanowsky2012}, \citet{Pogge2000} found little presence of nuclear dust which may suggest that no episodes of recent star formation have taken place. This suggests that the merger was probably dry.

 
\subsection{Age and metallicity model dependent estimations} 
In the following section, we compare the SEDs of M87 GCs with single stellar populations (SSPs) to describe the properties of the clusters in a more quantitative way. The same methodology is used in \citet{Montes2014}.

\subsubsection{Methodology}\label{modelfit}
In this section, the observed SEDs are compared with that of SSP models to obtain the properties of the stellar populations of the GCs, i.e.\ metallicity and age. Since our data are integrated luminosities, we convolved the theoretical SED spectra with the transmission curves of the photometric filters to retrieve synthetic photometry for comparison. The computed magnitude, in the AB system, for the $i^{th}$ filter is

\begin{equation}
 m^{i}_{AB}=-2.5\log\frac{\int_{\nu}F_{\nu}\phi^{i}(\nu)d\nu}{\int_{\nu}\phi^{i}(\nu)d\nu}+8.906,
\end{equation}

\noindent where $F_{\nu}$ is the theoretical SSP SED expressed in Jy, which is a function of metallicity, age and luminosity, and $\phi^{i}$ is the response curve of the $i^{th}$ filter. In order to find the most suitable SSP model, a reduced-$\chi^2$ minimization approach is applied.

\begin{equation}
   \chi^2=\frac{1}{N-n-1}\sum_{i=1}^{N}\frac{(m_{obs,i}-m_{mod,i})^2}{\sigma_i^2},
\end{equation}

\noindent where $m_{mod,i}$ depend on age, metallicity and luminosity, $N$ is the number of photometric filters, $n$ is the number of fitted parameters and $\sigma_i$ are the observational errors in the photometry of each band. In our case, $m_{mod,i}$ are the bias-corrected magnitudes while $\sigma_i$ are the errors derived from the simulations. As there are three parameters for the minimization (age, metallicity and luminosity), there will be at least five spectral bands required for the fit. The models used in this work are the Charlot \& Bruzual (2007, hereafter CB) models. The CB SSP models contain $221$ spectra describing the spectral evolution of SSPs from $0.1$ Myr to $20$ Gyr for $6$ different metallicities: Z=$0.0001$, $0.0004$, $0.004$, $0.008$, $0.02$ $(Z_{\odot})$, $0.05$. These models cover a range of wavelengths from $91$ \AA{} to $160$ $\mu m$. As the age--metallicity grid is irregular, the metallicity vector was resampled. The grid was expanded with $200$ metallicities linearly interpolated from the original SSPs. 

None of the SSP models used here is able simultaneously to fit the optical--NIR data together with the UV fluxes. This disagreement originates from the lack of EHB stars in these models. Consequently, fluxes at wavelengths shorter than $3000$ \AA{}, if available, were not used in our fits.

\subsubsection{Robustness of our results}\label{degeneracy}

In order to better assess the obtained results for the GCs of M$87$, we have first checked the reliability of our methodology. Below, we discuss the tests implemented to validate the ages and metallicities obtained from our SED fitting.

\paragraph*{Test with Galactic Globular Clusters:}

Testing our procedure with an observed template of known age and metallicity will enable us to confirm that our methodology accurately retrieves the properties of the clusters. The data of the Galactic globulars encompass the same wavelength sampling as  our clusters, but the determinations of age and metallicity have been done using other approaches, such as integrated spectroscopy and isochrone fitting of the main sequence turn-off. 

\begin{figure}
 \begin{center}
   \includegraphics[height=0.4\textheight]{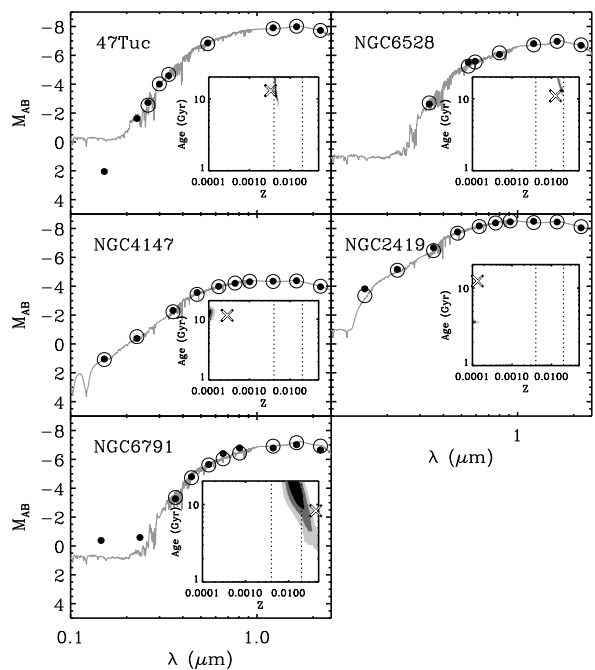} 
    \caption{Best fit models for the GGC SEDs. The filled circles represent the photometry of the clusters. The open circles indicate the convolution of the model with the filter response in each bandpass and the solid grey line is the best fit spectra. The insets show the corresponding age and metallicity maps with the $25$ (black), $68$ (dark grey) and $95$ (light grey) per cent confidence intervals. Crosses indicate the position in the maps of the values of age and metallicity of the clusters found in the literature. The dotted vertical lines indicate the range where STELIB models are reliable.}\label{figfitGC}
 \end{center}
\end{figure}

In this case, we convolved with the corresponding transmission curves (see Appendix \ref{sedGC} for more details on the photometry). In Fig.~\ref{figfitGC}, the best fits for each of the five sample clusters (Figure \ref{iplussed}) are shown. The insets show the age and metallicity confidence maps. The errors in age and metallicity are given by the shaded regions that contain the $25$, $68$ and $95$ per cent confidence levels. A comparison between literature values and the best fit is provided in Table \ref{tablafitmw}.

\begin{center}
 \begin{table}\caption{List of the literature values and best fits using our SED fitting tools found for the GGCs. The errors of the best fits represent the $1\sigma$ ($68$ per cent) confidence interval. The detailed information about the \textquotedblleft Literature\textquotedblright \ ages and metallicites of the clusters are given in Appendix B.}
  \begin{tabular}{|c|cc|cc|} \hline

      &  \multicolumn{2}{c|}{Literature}        & \multicolumn{2}{c|}{Best fit} \\
      & Age (Gyr) & \multicolumn{1}{c|}{[Fe/H]} & \multicolumn{1}{c}{Age (Gyr)} & [Fe/H] \\ \hline

\scriptsize{$47$ Tuc}   &  \scriptsize{$13.0\pm2.0$} & \scriptsize{$-0.70\pm0.1$} &  \scriptsize{$12.5^{+7.5}_{-3.2}$}     & \scriptsize{$-0.57^{+0.05}_{-0.03}$} \\ \hline
\scriptsize{NGC~$6528$} & \scriptsize{$11.0\pm1.0$} &  \scriptsize{$-0.1\pm0.2$}    & \scriptsize{$12.8^{+7.2}_{-0.6}$}     & \scriptsize{$0.08^{+0.03}_{-0.03}$} \\ \hline
\scriptsize{NGC~$4147$}  & \scriptsize{$11.5\pm1.0$} &  \scriptsize{$-1.8\pm0.3$}   & \scriptsize{$13.8^{+1.7}_{-4.5}$}     & \scriptsize{$-2.24^{+0.1}$} \\ \hline 
\scriptsize{NGC~$2419$}  & \scriptsize{$12.3\pm1.0$} &  \scriptsize{$-2.14\pm0.15$} & \scriptsize{$3.8^{+0.3}_{-0.1}$}      & \scriptsize{$-2.24^{+0.1}$} \\ \hline
\scriptsize{NGC~$6791$}  & \scriptsize{$8.3\pm0.5$}  &  \scriptsize{$0.45\pm0.04$}  & \scriptsize{$20.0^{+0.0}_{-4.0}$}     & \scriptsize{$-0.11^{0.3}_{-0.4}$} \\ \hline

  \end{tabular}\label{tablafitmw} 
 \end{table}
\end{center}

In the case of $47$ Tucanae, the UV bands were excluded from the fit since this cluster has a very red HB and radiates weakly in the FUV. It appears as one of the reddest cluster in UV colours in the study of \citet{Dorman1995}. Nonetheless, we find a very good agreement with the age and metallicity for this GC. 
Regarding NGC~$6528$, the best fit for this cluster recovers its age and metallicity accurately despite the lack of UV data due to constrains in GALEX detectors caused by the proximity of the cluster to the Galactic centre.
For NGC~$4147$, a difference between the literature and the recovered metallicity is found.  The fit of NGC~$2419$ yields an intermediate age for this cluster. The disagreement between the derived ages and metallicities of the latter clusters is probably caused by a bias towards solar metallicity due to the limitations of the stellar library of the CB models, STELIB. The STELIB library contains mostly stars in the $0.2\,Z_{\odot}<Z<Z_{\odot}$ range (see fig. 2 in \citealt{Vazdekis2010}, higher than the expected metallicity for NGC~$4147$). Therefore, the fit to these models tend to produce lower metallicities and/or young ages to compensate for this effect.
Apart from the disagreement due to the models, the result for NGC~$2419$ could also arise due to the presence of a significant ($\sim$30 per cent) second generation of stars with extreme helium enhacement \citep{DiCriscienzo2011}. \citet{Perina2011} showed that, using integrated photometry and/or spectroscopy, old M31 GCs can appear to be intermediate-age clusters due to the presence of hot HB stars. These stars contribute to enhance the UV flux as well as the strength of the Balmer lines mimicking young ages. Not including the UV bands in the fit of this cluster leads to the same result: the recovered age is younger than expected.
We also performed the fit for the open cluster NGC~$6791$. This cluster is known to host a significant population of EHB stars, consequently we excluded its UV data from the fit. Note that this fit does not reproduce the photometry in the $R$ and $I$ bands.
We can see that our SSP fits recover quite well the age and metallicity of the GGCs in the range of metallicities compatible with our clusters. \\

\paragraph*{Degeneracy due to photometric errors:}

Large photometric errors increase the uncertainty in the determination of ages and metallicities. To estimate the effects of degeneracy on age and metallicity due to these errors, we have carried out Monte Carlo simulations introducing different errors into SSP models with fixed age and metallicity. Our test consists of 10\,000 realizations of each model SED. Different error values were explored and gaussian noise was added to the photometry of the simulated clusters. For simplicity, we used the median metallicity derived in Section \ref{VIcolour}: Z=$0.004$ ([Fe/H]$=-0.7$). Two different ages were explored, $1$ and $10$ Gyr and errors of $0.05$, $0.1$ and $0.2$ mag were assumed. The age and metallicity distributions for the best fit models were analysed and compared to the actual values of the input models. 

As stated above, the age--metallicity degeneracy increases as errors increase. For a $10$ Gyr SSP, large errors produce a broad peaked histogram of recovered values. For median errors of $0.1$ mag, we recover the input values ($t>6$ Gyr and $0.0026>Z>0.0055$ dex) in $94$ per cent of cases, whereas for median errors of $0.2$ mag this lowers to $75$ per cent. In contrast, for a $1$ Gyr SSP and the same metallicity as above, the recovered ages are much more constrained around the input value (median $=1.0$ Gyr and $\sigma=0.4$ Gyr). However, metallicities tend to be richer. Consequently, to limit the possible degeneracy, we will select clusters with median photometric errors less than $0.1$ mag. This restriction ensures that we properly recover the ages and metallicities of $\sim$90 per cent of the clusters.

\paragraph*{Degeneracy due to limited wavelength range:}

The inclusion of UV and IR information is crucial to constrain the age and metallicity of stellar populations. \citet{Anders2004} concluded that a well sampled wavelength range, ideally including the NIR and UV, allows one to obtain reasonable SSP estimates. It was also stated that adding NIR data bands restricts the metallicity range. We tested the effect of including the IR fluxes. In this case, the photometry of the simulated clusters ranges from F$336$W to $K_s$. 

A model with $10$ Gyr and metallicity $Z=0.004$ becomes younger and more metal-rich when not including the UV. If only the optical is used, and restricting to errors below $0.1$ mag, only $47$ per cent of the total reproduces ages older than $6$ Gyr and metallicities in the range: $0.0026>Z>0.0055$ dex. If the $J$ band is included then the rate increases to $63$ per cent. Finally, if the $K_s$ band is added the rate increases to $86$ per cent. For the $1$ Gyr and $Z=0.004$ model the results present a growing dispersion around the actual values and do not show any degeneracy due to the limited wavelength sampling. This test proves the importance of including NIR photometry when studying old populations.

\paragraph*{Model dependency:}

It is beyond the scope of this paper to test the different SSP models and investigate the nature of the differences between them. However, in order to provide a qualitative overview of how different synthesis population models fit the SEDs, we compare our results with other SSP models. A new version of the \citet{Vazdekis2010} SSP models contains spectra for $50$ ages in the range from $63$ Myr to $17.78$ Gyr and for seven different metallicities: Z=$0.0001$, $0.0004$, $0.001$, $0.004$, $0.008$, $0.019$, $0.03$, spanning from  $3464$ \AA{} to $9468$ \AA{} (\citealt{Vazdekis2012}, hereafter V12; \citealt{Ricciardelli2012}). The CB models are based on a Chabrier IMF while the V12 models are based on the nearly identical Kroupa IMF.
Unfortunately, V12 models cover only the optical wavelengths, i.e.\ from F$410$M to F$702$W. In order to compare both sets of models consistently we have also estimated the age and metallicity in the optical using the CB models. The comparison was possible for $42$ clusters. Ages correlate linearly, although V12 model fits tend to produce an offset towards older ages. The agreement between the two models covers $66$ per cent of cases, although there is a bias towards high metallicities due to the limitations of wavelength sampling.

To summarize, IR data and a good signal-to-noise ratio are crucial to minimizing errors and breaking the age--metallicity degeneracy. It is expected that, in the future, models will include blue or extreme HB stars improving the determination of ages and metallicities using SED fits. Note that the bias introduced by the use of the STELIB library probably does not affect the results for our M87 GCs as the STELIB library contains mostly stars in the $0.004<Z<0.02$ range and the mean metallicity of our clusters is $Z\sim0.004$.

\subsubsection{Results}\label{results}

\begin{figure}
 \begin{center}
   \includegraphics[height=0.38\textheight]{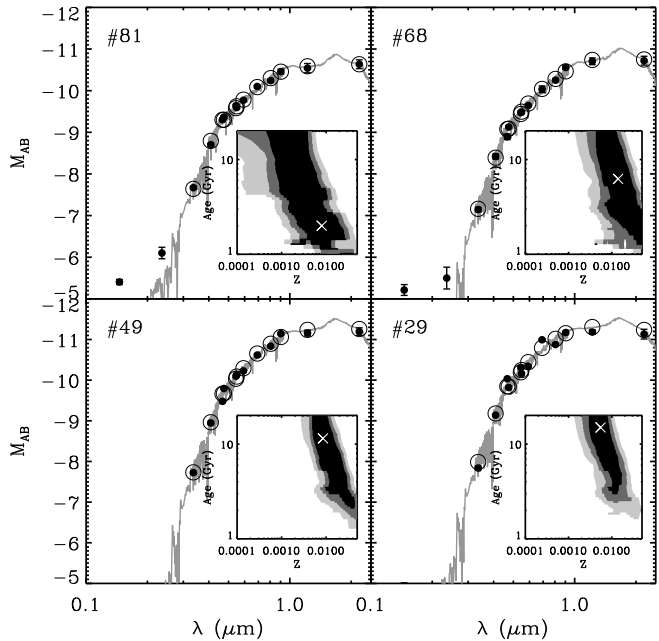} 
    \caption{SED of four of our clusters (black dots) and best fit CB model (grey line). Open circles represent the convolution of the model with the transmission curves for each filter. The insets show the age--metallicity parameter space and the contours for the $25$ (black), $68$ (dark grey) and the $95$ (light grey) per cent confidence intervals. }\label{figfit}
 \end{center}
\end{figure}

As shown in subsection \ref{degeneracy}, age and metallicity can be retrieved accurately for $\sim$90 per cent of cases when photometric errors are below $0.1$ mag and the NIR is present. For this reason, we restrict our analysis to a total of $13$ GCs.
Figure \ref{figfit} shows the best fit CB model for four clusters. The open circles represent the expected SEDs for the best fit model, while the black filled circles are the observed SEDs of the clusters. In the insets, the age--metallicity confidence maps are shown, with the black, dark grey and light-grey areas being the $25$, $68$ and $95$ per cent confidence intervals respectively. Note that the age--metallicity degeneracy associated with broadband photometry can be appreciated in the confidence maps.

 \begin{figure}
  \begin{center}
    \includegraphics[height=0.3\textheight, width=0.5\textwidth]{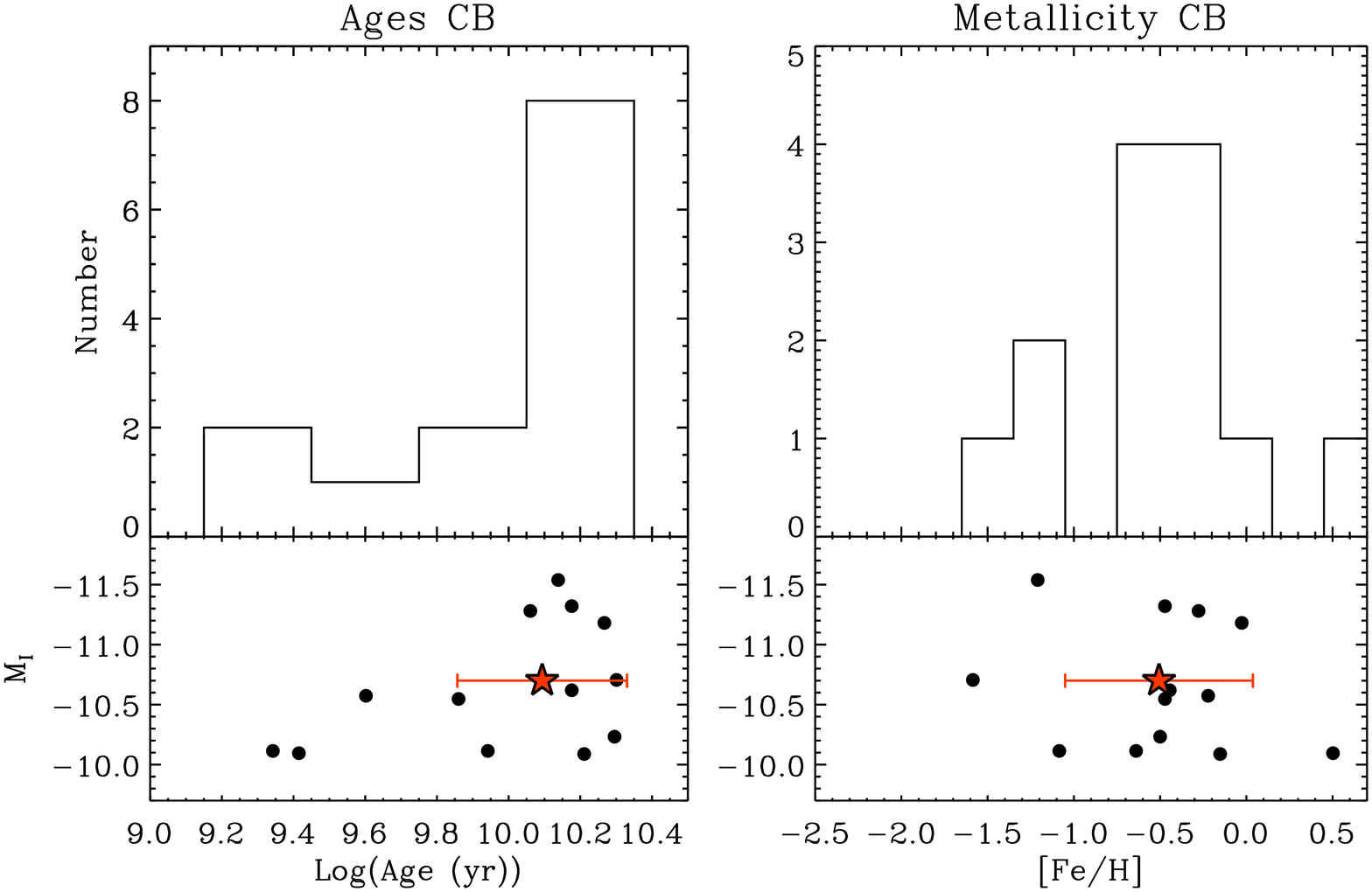} 
     \caption{Distribution of ages and metallicities for the $13$ GCs with NIR photometry. In each panel, the red star corresponds to the mean age and metallicity, respectively.}\label{fighist}
  \end{center}
 \end{figure}
 
To illustrate our results, Fig.\ \ref{fighist} shows the metallicity and age distributions of the best fits versus the absolute magnitude in $I$ ($M_{I}$, Vega system) for our clusters with IR photometry. In Figure \ref{fighist}, we also indicate the mean age and metallicity and their corresponding uncertainty. The mean parameters are: $\log (\rm{Age(yr)})=10.1\pm0.2$ and [Fe/H]$=-0.5\pm0.5$ dex. The distribution of metallicities seems compatible with the broad but single-peaked distribution derived in Section \ref{VIcolour}, although a more complicated distribution (e.g. double-peaked) cannot be rejected owing to the limited number of clusters. The metallicities derived are in good agreement with those derived independently in Section \ref{compGGCs} and Section \ref{VIcolour}. Note that for three ($20$ per cent) of the clusters we retrieve young ages. It is very likely that these young ages are not real but are caused by the age--metallicity degeneracy (two of these clusters lack $K_s$ band photometry and the errors in the $K_{\rm s}$ band are large in the other cluster: $0.20$ mag). This fact emphasizes that the $K_s$ band is crucial for breaking the mentioned degeneracy.

\section{Discussion}\label{discussion}

The M87 GC system is one of the most extensively studied GC systems. In our study we have dealt with the innermost GCs, within $\sim3\,\times\,3$ kpc$^2$, using a broad wavelength sampling from the UV to the NIR. This is the first multiband analysis of these inner M87 clusters that includes the NIR range at subarsecond resolution. As the centres of elliptical galaxies are the relics of the early stages and the GCs are formed in intense episodes of star formation, these clusters can help us to constrain models of the formation of M87.

In this section, we discuss the properties derived from our analysis compared to the whole population of GCs in M87.

\subsubsection*{Is the GC metallicity distribution bimodal?}

As mentioned in Section \ref{Intro}, the majority of the studies carried out on extragalactic GCs are limited to the optical range, making colour bimodality one of the most commonly observed properties of the clusters \cite[see][and references therein]{Harris2009}. In the case of M87, the blue and red clusters show peaks at $V-I\sim 0.95$ and $1.20$, respectively \citep{Kundu1999, Peng2009}. 
The mean value of the colour distribution of our GC sample is $V-I = 1.1$, which is compatible with the mean colour in \citet{Kundu1999, Kundu2007} and \citet{Peng2009}. We found no clear evidence of colour bimodality in our data (see Sec.\ \ref{VIcolour}). This might be produced by the small number statistics in our sample or, alternatively, by a selection bias: red clusters may be the dominant population in the circumnuclear region of M87 \citep{Harris2009}. \citet{Strader2011} derived the surface density profiles of both subpopulations of GCs and found that red GCs dominate in the inner $2$ arcmin. Consequently, most of the GCs in our sample are red GCs.

The physical meaning of this bimodality remains controversial although the most commonly accepted interpretation is that the colour distribution translates into a metallicity bimodal distribution \citep[see][]{Brodie2006}. However, \citet{ChiesSantos2011b, ChiesSantos2011a} found that the colour bimodality disappears when IR colours ($g-K$) are considered (see their figure 9). Furthermore, \citet{ChiesSantos2012} concluded that a bimodal optical colour distribution does not necessarily imply a bimodal metallicity distribution, but is a consequence of a non-linear colour--metallicity conversion (see also, \citealt{Yoon2006, Peng2006}). Metallicities derived photometrically produce [Fe/H] $\simeq-1.5$ dex for the blue clusters and [Fe/H] $\simeq-0.3$ dex for the red clusters \citep{Kundu1999, Jordan2002}. The metallicities derived by \citet{Kaviraj2007}, based on SSP fits from the UV to $I$, also agree with these results. In the same direction, \citet{Yoon2011} found that the metallicity distribution function of M87 GCs peaks at [Fe/H] $\simeq-0.5$. The mean metallicity of our sample of GCs is $<$ [Fe/H] $>\simeq-0.6$ (Z$\simeq0.004$, the mean of the different estimates derived throughout this paper), compatible with the values reported in previous studies. This places our sample in the red (metal-rich) subpopulation, although we  also include some blue (metal-poor) GCs. These blue GCs could be also located in the outskirts of M87 but are seen in its central region due to a projection effect.

Based on spectroscopy, \citet{Cohen1998} derived ages and metallicities for $150$ M87 GCs. They found a marginally bimodal metallicity distribution with a mean metallicity of $<$[Fe/H]$>=-0.95\pm 0.5$. This value is compatible but slightly lower than the results found here using SSP fitting ([Fe/H ]$=-0.5\pm0.5$). This is also caused by the location of their GCs at larger radii (see above, \citealt{Strader2011}). Unfortunately, all their clusters are beyond our field of view and direct comparison is not possible.

Regarding ages, \citet{Kundu1999}, using $V$ and $I$ \emph{HST}/WFPC2 colours, found an age difference between the two subpopulations. They estimated an age of $15$ Gyr for the blue cluster subpopulation and $9$--$12$ Gyr for the red ones. However, subsequent studies proved them to be equally old to within errors \citep[e.g.][]{Jordan2002} and the difference in colours can only be explained by a difference in metallicities. \citet{Cohen1998} also found that the spectrocopic ages of the $150$ GCs are compatible with $\sim$13 Gyr. The ages obtained from our SSP model fitting are not well constrained due to the age--metallicity degeneracy, although the resulting ages concentrate around $12$ Gyr with no evidence of two subpopulations in age.

Summarizing, according to our results the central GCs of M87 are old and with a mean metallicity of [Fe/H] $\simeq-0.6$ dex, the richest within the population of M87 GCs. We will discuss later on the implications of this for understanding the formation of the innermost region of M87.

\subsubsection*{GCs vs spheroidal stellar population}

The GCs of M87 seem to be closely linked to the formation history of the galaxy. The ages of the halo stars and the GCs are similar \citep{Kuntschner2010, Cohen1998}. In addition, the spatial distribution of the red subpopulation of M87 GCs in the inner regions of the galaxy, as studied by \citet{Kundu1999} and more recently by \citet{Harris2009}, mimics fairly closely the metal-rich spheroid light of the host galaxy \citep{Strader2011, Forte2012}.  
\citet{Kuntschner2010} studied M87 spectroscopically  using SAURON, and determined that the age of its stellar population is old ($>10$ Gyr) and its metallicity is supersolar (Z~$\sim0.03$) in the inner parts of the galaxy ($R = R_e/8$, similar to our field of view). 
The mean metallicity of our GC sample is [Fe/H] $\simeq -0.6$, which is about eight times lower than that of the galaxy itself. Similar results were found by \citet{Jordan2004} in a study of four cD galaxies (imaged with the WFPC2), where an offset of $\Delta$[Fe/H] $\sim0.8$ dex was found between the GCs and the stellar population of their host galaxies. Consequently, a metallicity offset towards lower metallicities seems a natural outcome if the chemical enrichment process of M87 GCs ceased earlier than that the bulk of stars of the galaxy, as discussed in \citet{Yoon2011}. 

In this sense, we can sketch the following picture for the formation of M87. Halo stars and GCs were formed in the early phase of the collapse of the gas cloud. Later on, as the gas collapse continued, an intense burst created the nucleus of M87  \citep[see ][ for a longer discussion]{Montes2014}. That phase took place around $12$ Gyr ago. Subsequent mergers with other galaxies could have brought both metal-poor GCs and the stars of the outer evelope of M87. The fact that no young GCs have been observed in M87 suggests that this merging process was basically dry (see below).

\subsubsection*{Evidence of a dry merger}

Examples of the dry merging process discussed above could have been detected observationally. In fact, the detection of dust lanes in the circumnuclear region of M87 \citep{Pogge2000} implies that a merger event took place after the galaxy formation, which does not seem to have induced episodes of formation of GCs, as shown in Section \ref{Halpha}. The presence of these lanes suggests that the merger occurred $\lesssim1$ Gyr ago \citep{Rudick2009, Romanowsky2012}. Actually, age estimates obtained from comparison with Galactic GC and SSP fitting indicates that the GCs studied are old ($>10$ Gyr). However, three of the clusters exhibit rather young ages (1--4 Gyr), probably caused by the methodology applied.

These facts, together with the evidence that a low-luminosity active galactic nucleus is present in M87, suggest that whatever merger activity occurred in the past it was dry. An observation that is consistent with this idea is that the galaxies near the core of galaxy clusters are almost depleted of their gas content \citep[e.g.][]{Chung2009}. Another relevant finding is that of \citet{Romanowsky2012}, who show that a substructure in the position--velocity phase space is compatible with the expectations of the recent accretion of a dwarf galaxy into M87. 

\section{Summary and conclusions}

We have compiled multiwavelength photometric data for $110$ GCs detected in a region of $\sim\,3\times 3$ kpc$^2$. These data are based on high spatial resolution images obtained with VLT/NaCo and instruments on board \emph{HST}. The SEDs for these GCs were built covering a wide wavelength range, from the FUV to $K_s$. The use of adaptive optics techniques turns out to be crucial in order to detect and measure GCs close to the nucleus of M87. 

The M87 GC SEDs were compared to those of the Milky Way GCs and the mean SED of the M87 GCs is very similar to that of $47$~Tuc in the optical--IR passbands but not in the UV range. A similar UV excess (probably due to a hot HB population) as shown by the M87 GCs was only present in the SED of the open cluster NGC~6791. 

Furthermore, ages and metallicities were derived for $13$ M87 GCs after fitting SSP models. We performed simulations accounting for the relevance of the different wavelength sampling and photometric errors in the metallicity/age determinations. We conclude that about $90$ per cent of the GCs have valid ages and metallicities when photometric errors are below $0.1$ mag and the NIR passbands are included. The NIR (especially the $K_s$ band) has proven to be a useful tool for breaking the degeneracy and thus determining accurate parameters for extragalactic objects.

Most of the metallicities derived for our GCs distribute around the red (metal-rich) peak of the bimodal metallicity distribution, although there is no hint of bimodality in our data. A median value of $<$[Fe/H]$> \simeq -0.6$ (Z$\simeq 0.004$) was obtained, a result which is in good agreement with previous photometric determinations of the metallicity in the inner regions of M87. The resulting ages are compatible with an old population ($>10$ Gyr), which is consistent with previous spectroscopic determinations of the age of GCs in the outer parts of M87. These derived ages are also compatible with the ages of the oldest Galactic GCs. This indicates that no wet merging processes have occured recently. 

The average metallicity derived here for the innermost GC system is lower (Z$\simeq 0.004$) than that of its host galaxy (Z$\sim 0.03$). Taking into account that the spatial distribution of the clusters and that of the field star light are similar, this result suggests that the formation of GCs stopped before the formation of the bulk of stars. 

\section*{Acknowledgements}
We would like to thank A. Vazdekis and E. Ricciardelli for many useful comments on SSPs models, I. Trujillo for his comments on galaxy formation and evolution and M. Beasley for his comments on the paper. We also thank T. Mahoney (IAC) for the English edition of this manuscript. We also thank the anonymous referee for comments that improved the quality of the paper. This work is partially funded by the Spanish MEC project PAYA 2011-25527.

\bibliography{m87}
\bibliographystyle{mn2e}

\bsp
\appendix
\newpage
\section[]{M87 GCs photometric accuracy} \label{simulations}

\begin{figure*}
  \begin{center}
    \begin{tabular}{c}
    \includegraphics[height=0.29\textheight]{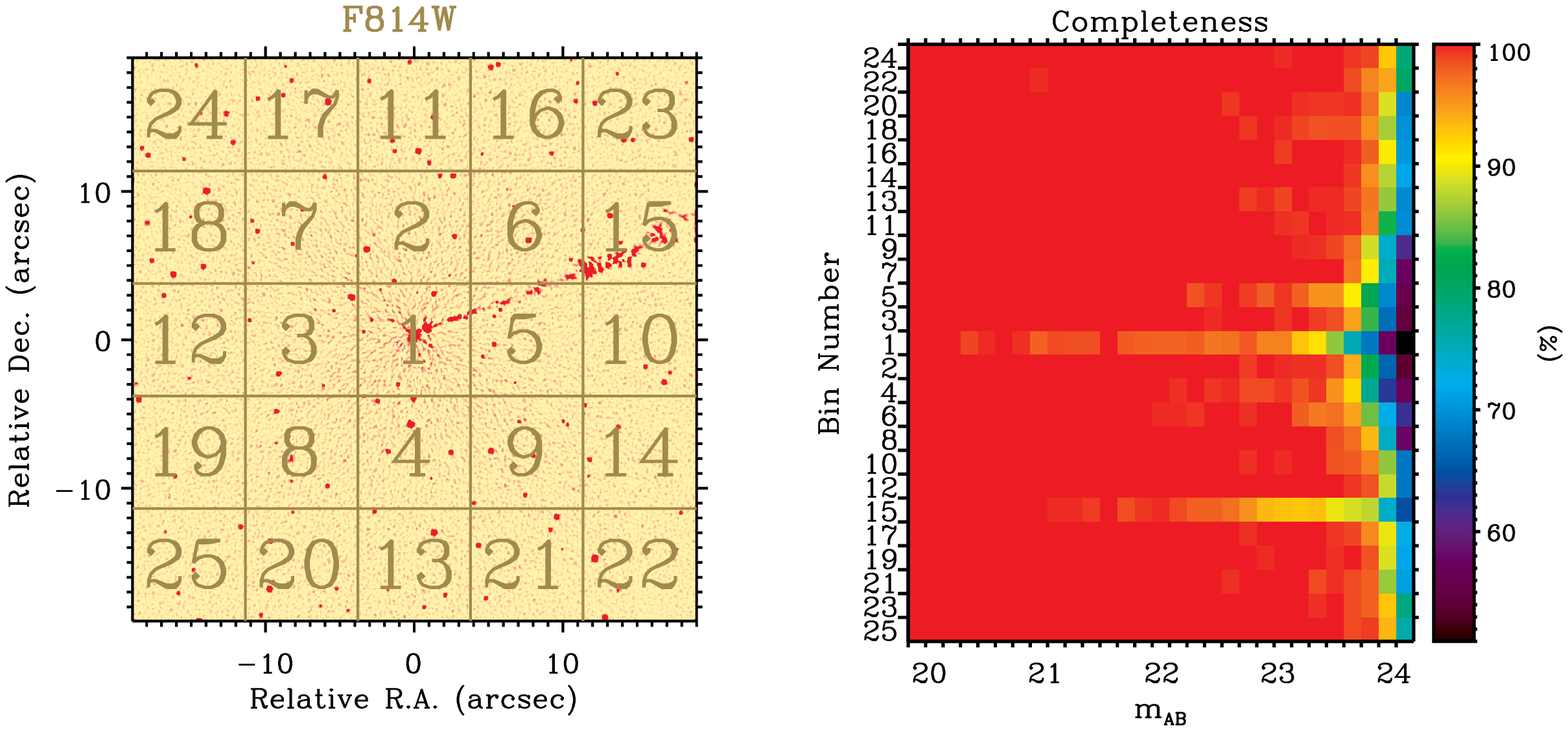} \\ 
    \includegraphics[height=0.29\textheight]{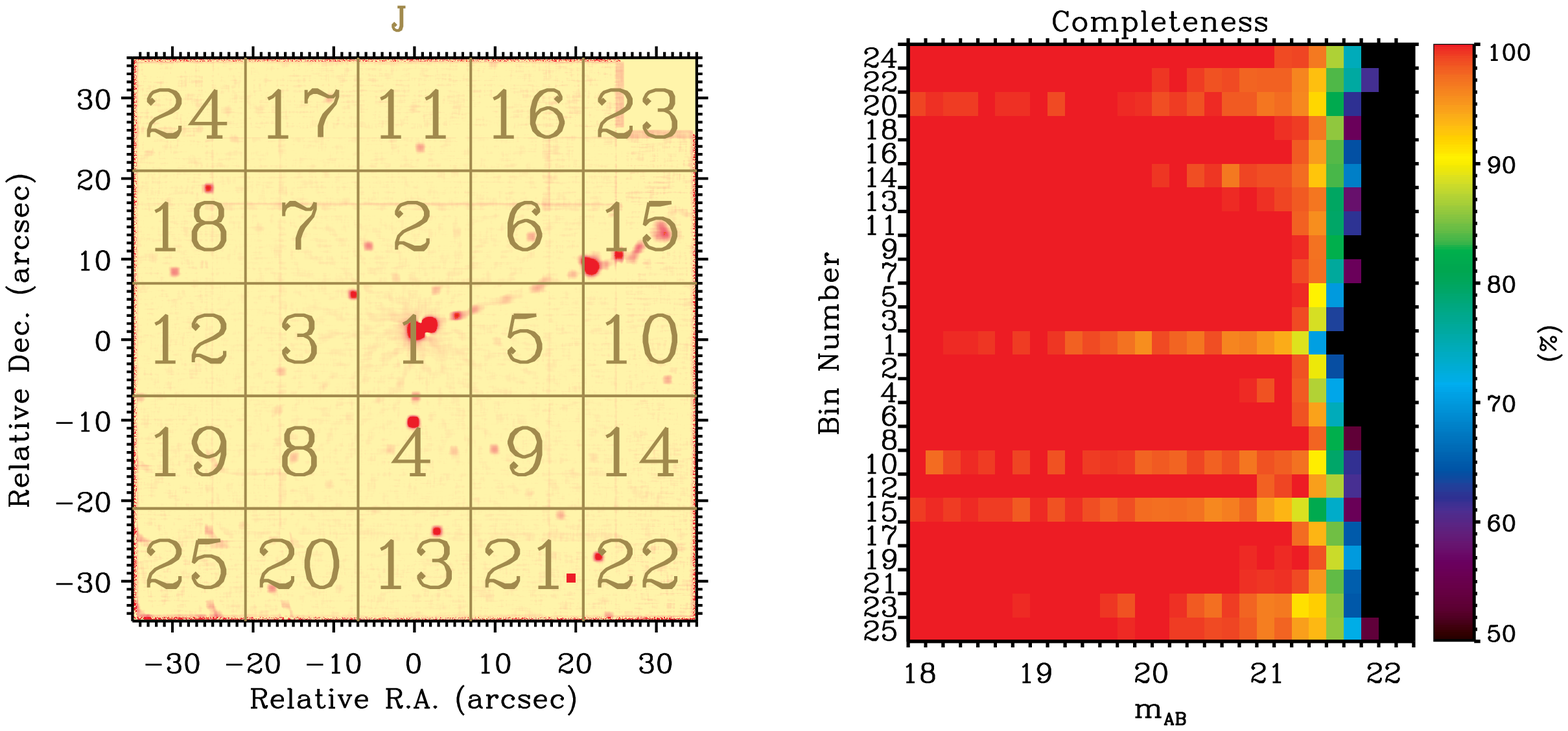}\\
    \includegraphics[height=0.29\textheight]{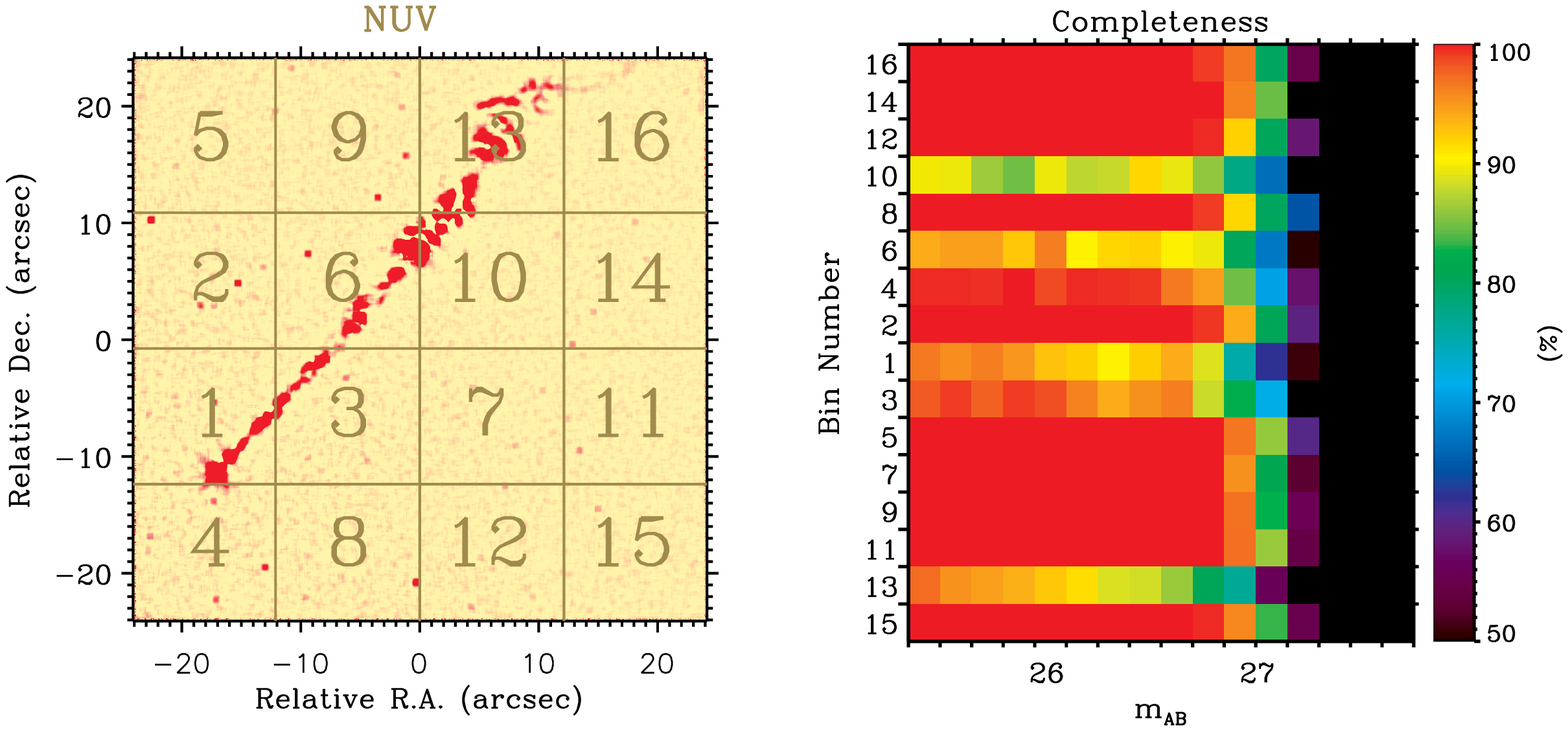} \\
    \end{tabular} \caption{The three panels represent the completeness derived from the simulations for three different bands; the NUV, the optical (F814W) and the IR (J). The left panels show the division in numbered bins of the image. Note that the NUV has a different image alignment and consequently a different bin configuration.}\label{simus}
 \end{center}
\end{figure*}

Getting reliable photometry of M87 GCs in the inner regions is not an easy task as the underlying galaxy outshines the light of these systems. This is particularly significant in the redder bands as the galaxy light contribution increases towards the NIR. 

To assess the accuracy of our photometry, we performed artificial cluster tests to determine the completeness and errors. To model the simulated clusters, a Moffat function was used. Realistic parameters of the Moffat function were retrieved by fitting this function to isolated GCs on the real images using \textsc{GALFIT} \citep{Peng2002}. The residuals of the fitting to the clusters were less than $20$ per cent in all cases.

The simulated clusters were randomly distributed across the image. The magnitudes of the simulated clusters expanded the range of the observed GCs. As the influence of the background light of M87 changes through the image, to calculate the parameters of our photometric corrections we have divided the distributions, both spatial and in magnitude, in bins of $6\farcs0$ and $0.15$ mag, respectively. The number of simulated clusters per bin was $250$. This gives us good statistics for estimating the corrections. 
The artificial clusters in all filters thus created were subjected to the same analysis as the real data. \\

\begin{figure}
  \begin{center}
   \includegraphics[scale=0.5]{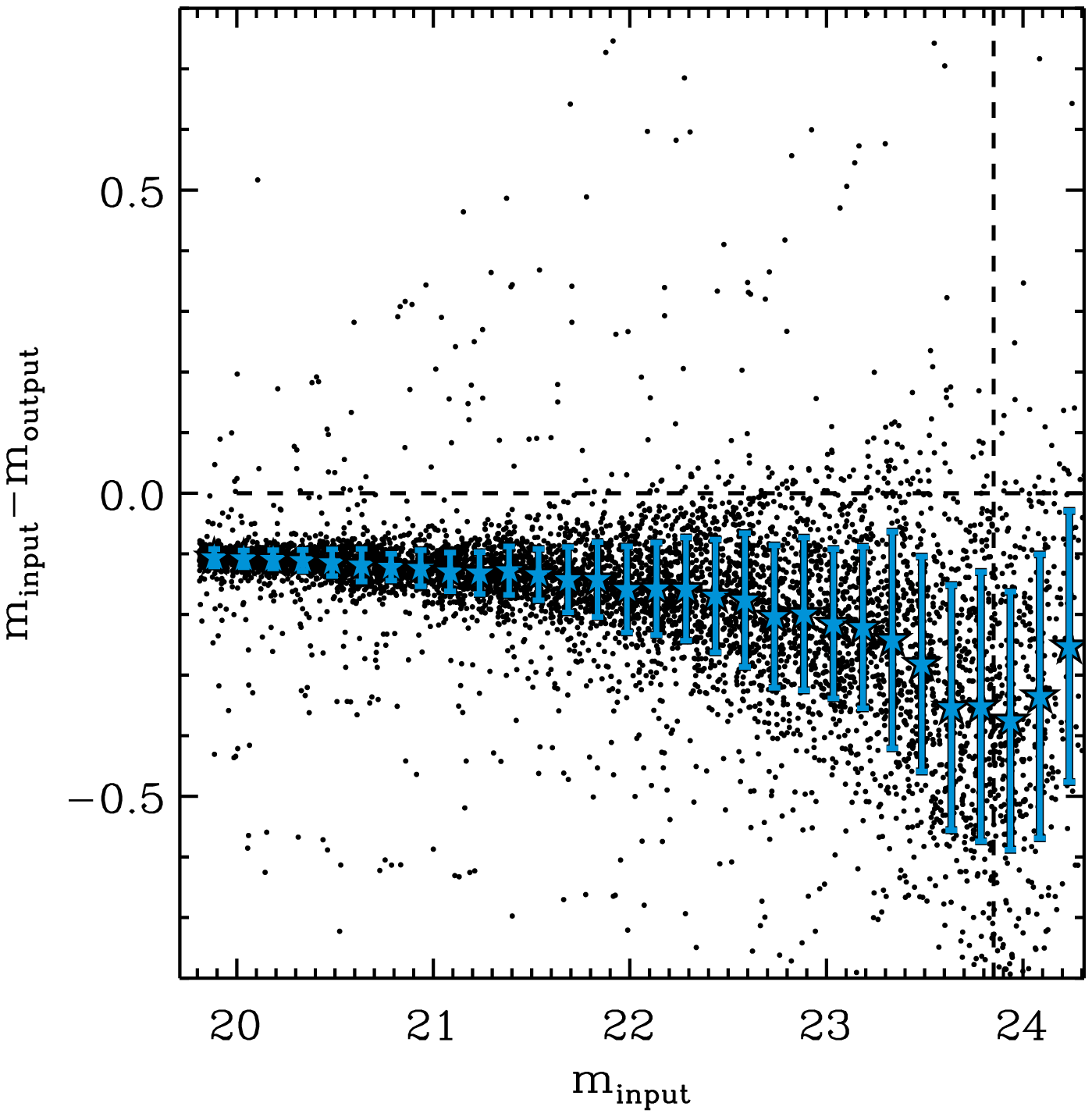} 
   \caption{Input magnitude vs Input--Output magnitude for one bin. This figure illustrates the bias and error for a given bin. The bias is the mean of the difference between the input and the recovered clusters in a bin of magnitude (blue stars) The blue error bars are the dispersion of the points around this mean and thus represent our simulated error. The vertical dashed line represents $90$ per cent completeness in this spatial bin.}\label{figbiaserr}
 \end{center}
\end{figure}

Completeness, bias and errors were derived to account for the reliability of the GC photometry depending on the position of the GCs in the images (i.e.\ different background contamination as a function of the distance to the centre of the galaxy). The completeness represents the rate of recovered simulated GCs per magnitude and spatial bin. The bias is the median of the difference between the input magnitudes and the recovered magnitudes. Finally, the error is the scatter of the difference between the input and output magnitudes. For a visual description of these parameters, Fig.~\ref{figbiaserr} represents the bias (blue stars) and the error (blue error bars) for a spatial bin. 

An example of three completeness maps for three filters covering the studied wavelengths in this work can be seen in Figure \ref{simus}. The bins are numbered, ordered with increasing distance to the galaxy nucleus, shown in the leftmost panel. As expected, the detection efficiency depends on the projected radius from the centre of the galaxy. The bias and error increase with decreasing distance, where the background light of the galaxy is more intense and also when the magnitudes of the GCs are fainter. The effect of the presence of the jet in bin \#$15$ is seen.

The completeness tends to be higher in filters with large exposure times, as the ACS/F$606$W and ACS/F$814$W and with a low contribution of the background galaxy light, as in the STIS UV bands. The bias is affected by the noise. Shorter exposure images have higher bias. Also, the bias is higher in redder filters as the emission of the diffuse light of the galaxy increases. 

\section[]{Galactic Globular/Open Clusters: Data}\label{sedGC}

Multiwavelength SEDs of the integrated light of GCs in our Milky Way galaxy do not exist. These SEDs are, however, very important for comparative extragalactic studies where only the broadband colours are accesible. 
In this paper Milky Way GCs $47$ Tucanae, NGC~$6528$, NGC~$4147$ and NGC~$2419$ were selected to construct UV-to-NIR light-integrated SEDs. The selection of clusters was mostly driven by the availability of wide field images over a wide wavelength range. They were also chosen to cover a wide range of metallicities ($-2.14<$ [Fe/H] $<-0.1$). For each cluster, 2MASS, \emph{HST}, SDSS and GALEX images were used. We performed aperture photometry within an aperture radius equal to the core radius of the clusters taken from the Harris catalogue \citep{Harris1996}. The centre of the aperture was set on the centre of the cluster, as defined in \citet{Goldsbury2010} or, alternatively, in the Harris catalogue. In the case of \emph{HST} imaging, short exposure frames were chosen to avoid saturation due to bright stars. The main source of error in the integrated magnitudes is the  determination of the photometric zero point, as Poisson noise is too small due to the high photon counts we are integrating.
We also added from the literature the open cluster NGC~$6791$, for its UV properties with respect to its high metallicity such as is not found in any globular cluster in the Milky Way.

\subsubsection*{47 Tucanae}
This  is the second largest and brightest globular cluster in the sky and for this reason it is one of the most studied old associations of the Milky Way. Its age and metallicity have been measured both spectroscopically \citep[e.g.][]{Schiavon2002, Mcwilliam2008} and photometrically \citep[i.e.][]{Salaris2007} and both determinations are in good agreement. The adopted values for the metallicity, age and extinction are: [Fe/H] $=-0.7\pm0.1$, $13.0\pm2.0$ Gyr and $E(B-V)=0.04$. To built its SED, we used FUV and NUV images from GALEX\footnote{Galaxy Evolution Explorer, see http://www.galex.caltech.edu/}, F$336$W, F$255$W, F$300$W and F$555$W from \emph{HST} and $J$, $H$ and $K_s$ from 2MASS\footnote{Two-Micron All-Sky Survey, see http://www.ipac.caltech.edu/2mass/}.

\subsubsection*{NGC~6528}
This globular cluster is located in Baade's Window and its relatively low extinction value allows us to use it as a template of an old, metal-rich, bulge globular cluster. Its location near the Galactic centre makes it a natural choice for comparison with our sample of central M$87$ GCs. The values of metallicity, age and extinction are: [Fe/H] $=-0.1 \pm 0.2$ \citep{Zoccali2004}, $11.0 \pm 2.0 $ Gyr \citep{Feltzing2002} and $E(B-V)=0.55$ \citep{Momany2003}. We used the following filters: F$336$W, F$555$W,$F606$W and F$814$W from \emph{HST} and $J$, $H$ and $K_s$ from 2MASS. UV data were not found for this cluster.

\subsubsection*{NGC~4147}
This cluster is located in the halo of the MW. NGC 4147 is intrinsically rather small and  few detailed studies of it are available. It has an age of  $11.5 \pm 1.0$ Gyr \citep{Marin-Franch2009} and [Fe/H] $=-1.8\pm0.3$ \citep{Rutledge1997} and $E(B-V)=0.026$ from \citet{Schlegel1998}. We use FUV and NUV images from GALEX, $u'$, $g'$, $r'$, $i'$ and $z'$ images from SDSS\footnote{Sloan Digital Sky Survey, see http://www.sdss.org/} and $J$, $H$ and $K_s$ images from 2MASS.

\subsubsection*{NGC~2419}
NGC 2419 is a very bright, unusually large GC located at a very large distance from the centre of the Milky Way. It is also one of the most massive MW GCs. It has shown evidence for two subpopulations of stars. It also has a strong EHB \citep[see][]{DiCriscienzo2011}. Its properties are: [Fe/H] $=-2.14\pm0.15$\citep{Harris1997}, $E(B-V)=0.11$ \citep{Harris1996} and $12.3\pm1.0$ Gyr \citep{Salaris2002}. FUV and NUV are from GALEX, $u'$, $g'$, $r'$, $i'$ and $z'$ from SDSS and $J$, $H$ and $K_s$ from 2MASS.

\subsubsection*{NGC~6791}
This is an $8.2\pm0.5$ Gyr \citep{Grundahl2008} massive open cluster with supersolar [Fe/H] $=+0.45\pm0.04$ \citep{Anthony-Twarog2007} metallicity. It has hot sources identified as EHB stars \citep[see][and references therein]{Buzzoni2012}. The photometry, from the UV to the NIR, was taken from \citet{Buzzoni2012}. As \citet{Buzzoni2012} do not include errors for their photometry we assumed a relative error of $2$ per cent.

In Figure \ref{figgalgc} the SEDs for the galactic clusters are shown. The photometric data can be found in Table \ref{tablamw}.

\onecolumn
 \begin{landscape}
  \begin{center}
  \begin{table}
\footnotesize

  \begin{tabular}{c|cccccccccc}
\rowcolor{LightGray}
47 Tuc & &FUV  & NUV  & F255W & F300W  &F336W  &F555W  & J & H & K\\ \hline
 &    & $15.4 $ & $11.7 $ & $10.6$ & $9.4 $&$ 8.8$ & $6.51 $ & $5.46$ & $5.39$ & $5.64 $ \\
 &    &$\pm 1.6$ & $ \pm 0.5$ & $\pm 0.4$ & $\pm 0.2 $&$ \pm 0.2$ & $ \pm 0.07$ & $ \pm 0.04$ & $\pm 0.04$ & $\pm 0.05$ \\ 
\rowcolor{LightGray}
NGC 6528 & & & &F336W  & F555W  & F606W & F814W  & J & H & K \\\hline
 & & & &$13.5 $ & $10.6 $ & $10.5 $  & $10.1$  & $9.3 $  & $9.2  $ & $9.5 $ \\ 
 & & & &$\pm 0.3$ & $ \pm 0.1 $ & $\pm 0.1$  & $\pm 0.1$  & $\pm0.1$  & $ \pm 0.1 $ & $ \pm0.1$ \\ 
\rowcolor{LightGray}
NGC 4147 & FUV & NUV  & u & g & r & i & z & J & H & K \\ \hline
 &$17.5 $& $16.1  $&$14.1  $&$12.92 $&$12.49 $ &$12.28  $&$12.18  $&$12.14$ &$12.10 $ &$12.53 $ \\ 
 &$\pm 0.4 $& $ \pm 0.2 $&$ \pm 0.1 $&$ \pm 0.06 $&$ \pm 0.05$ &$ \pm0.04 $&$ \pm 0.04 $&$ \pm 0.04$ &$ \pm 0.04$ &$\pm 0.06$ \\ 
\rowcolor{LightGray}
NGC 2419 & FUV & NUV & u & g & r & i & z & J & H & K \\\hline
  &$16.0 $ &$14.6 $ &$13.1 $ &$12.07 $ &$11.65 $ &$11.44 $ &$11.36 $ &$11.33$ &$11.37$ &$11.78 $ \\ 
  &$\pm 0.3$ &$\pm0.1$ &$\pm0.1$ &$\pm 0.08$ &$\pm0.07$ &$\pm0.06$ &$\pm0.06$ &$\pm0.06$ &$\pm0.07$ &$\pm0.08$ \\
\rowcolor{LightGray}
NGC6791 & FUV & NUV &  U & B & V & R & I & J & H & K \\\hline
  &$13.07$ & $12.86$ & $10.18$ & $8.65$ & $7.80$ & $7.04$ &
$6.66$ & $6.67$ &  $6.43$ & $6.79$\\ \hline
   \end{tabular}\caption{Data for the MW Globular/Open Clusters.}\label{tablamw}

\normalsize
 \end{table}
  \end{center}
 \end{landscape}

\twocolumn

\section[]{Data for M87 Globular Clusters}\label{datam87}

The multiwavelength photometric data for the $115$ M$87$ GCs are presented in Tables \ref{tablaradec}, \ref{tablacums} and \ref{tablacumsnobias}. Table \ref{tablaradec} presents the number of the clusters and their distance to the centre. Tables \ref{tablacums} and \ref{tablacumsnobias} contain the number assigned to the cluster in this work (first column), the radial distance relative to the centre (second column) and the magnitudes in the AB system for the $15$ filters used (column $2$-$15$) with their errors. The photometry is corrected for Galactic reddening and bias and the errors correspond to the errors derived in Sec.\ \ref{simulations} in \ref{tablacums}. In Table \ref{tablacumsnobias}, we did not correct the photometry bias or extinction, but the errors are from the simulations.

The objects removed from the analysis are: $\#54$ and $\#58$ (faint sources near the active nucleus and probably contaminated by it), $\#74$ (a compact and extremely red object in $V-I$, possibly a star), $\#82$ and $\#85$ (diffuse and red in $V-I$, probably background galaxies). 
The asterisks mark those objects whose magnitudes are fainter than $68$ per cent completeness in the simulations. Table \ref{tablaupp} shows the upper limits in magnitude for three radial distances from the nucleus of M$87$. The $13$ clusters used in Section \ref{results} are marked with a cross.

\onecolumn
\setlongtables

\begin{center}
\begin{longtable}{ccccc}
\caption{Coordinates of the detected $115$ sources relative to the nucleus of M87. The position of the nucleus is: RA=$ 12\rm{h}\,30\rm{}m\, 49.4311\rm{s}$, Dec=$12\rm{d}\,23\rm{m}\,29.1683\rm{s}$} \\
    \# & $\Delta RA$ (arcsec) & $\Delta Dec$ (arcsec) & \citep{Kundu1999} ID & \citep{Sohn2006} ID \\ \hline
   $   1   $&$   10.54 $&$  -18.96 $&$     --- $&$     ---  $\\
$   2   $&$  -13.12 $&$  -19.14 $&$     --- $&$     ---  $\\
$   3   $&$  -13.97 $&$   17.41 $&$     --- $&$     ---  $\\
$   4   $&$    8.44 $&$   16.95 $&$      85 $&$    3104  $\\
$   5   $&$    3.09 $&$   16.91 $&$      73 $&$    3101  $\\
$   6   $&$  -11.25 $&$   16.79 $&$     --- $&$     ---  $\\
\end{longtable} 
\end{center}

\begin{landscape}
 \begin{longtable}{cccccccccccccccc} 
  \caption{Photometrical data for the M$87$ Globular Clusters. Magnitudes are corrected using the simulations.} \\
   \# & FUV & NUV &  F336W & F410M & F467M & F475W & F555W & F547M & F606W & F658N & F702W & F814W & F850LP & $J$ & $K_s$\\
  &  \multicolumn{15}{c}{\cellcolor[gray]{0.8}[mag AB]}\\
   \hline
  \endfirsthead
   \caption{\emph{Continued}} \\
   \# &  FUV & NUV &  F336W & F410M & F467M & F475W & F555W & F547M & F606W & F658N & F702W & F814W & F850LP & $J$ & $K_s$\\
   &  \multicolumn{15}{c}{\cellcolor[gray]{0.8}[mag AB]}\\
   \hline
  \endhead 
   \hline
  \endfoot 
    $  1 $& $         .....  $ & $         .....  $ & $         26.18* $ & $          24.30 $ & $          23.25 $ & $          23.40 $ & $          22.64 $ & $          23.06 $ & $          23.00 $ & $         .....  $ & $          22.58 $ & $          22.60 $ & $         22.58* $ & $         <21.36 $ & $         .....   $ \\
 $    $& $                $ & $                $ & $       \pm 0.44 $ & $       \pm 0.14 $ & $       \pm 0.16 $ & $       \pm 0.05 $ & $       \pm 0.07 $ & $       \pm 0.10 $ & $       \pm 0.06 $ & $                $ & $       \pm 0.11 $ & $       \pm 0.10 $ & $       \pm 0.13 $ & $                $ & $                 $ \\
 $  2 $& $         .....  $ & $         .....  $ & $         .....  $ & $          22.80 $ & $          22.46 $ & $          22.07 $ & $          21.72 $ & $          21.80 $ & $          21.66 $ & $          21.21 $ & $         .....  $ & $          21.18 $ & $          20.98 $ & $         .....  $ & $         <21.01  $ \\
 $    $& $                $ & $                $ & $                $ & $       \pm 0.06 $ & $       \pm 0.06 $ & $       \pm 0.02 $ & $       \pm 0.03 $ & $       \pm 0.02 $ & $       \pm 0.02 $ & $       \pm 0.07 $ & $                $ & $       \pm 0.04 $ & $       \pm 0.04 $ & $                $ & $                 $ \\
 $  3 $& $         .....  $ & $         .....  $ & $         .....  $ & $         .....  $ & $         .....  $ & $         <24.18 $ & $         24.81* $ & $         .....  $ & $          24.18 $ & $         .....  $ & $         .....  $ & $          23.24 $ & $         <22.27 $ & $         <21.22 $ & $         .....   $ \\
 $    $& $                $ & $                $ & $                $ & $                $ & $                $ & $                $ & $       \pm 0.15 $ & $                $ & $       \pm 0.13 $ & $                $ & $                $ & $       \pm 0.15 $ & $                $ & $                $ & $                 $ \\
\end{longtable}
\end{landscape}

\begin{landscape}
 \begin{longtable}{cccccccccccccccc} 
  \caption{Photometrical data for the M$87$ Globular Clusters uncorrected from bias.} \\
   \# & FUV & NUV &  F336W & F410M & F467M & F475W & F555W & F547M & F606W & F658N & F702W & F814W & F850LP & $J$ & $K_s$\\
  &  \multicolumn{15}{c}{\cellcolor[gray]{0.8}[mag AB]}\\
   \hline
  \endfirsthead
   \caption{\emph{Continued}} \\
   \# &  FUV & NUV &  F336W & F410M & F467M & F475W & F555W & F547M & F606W & F658N & F702W & F814W & F850LP & $J$ & $K_s$\\
   &  \multicolumn{15}{c}{\cellcolor[gray]{0.8}[mag AB]}\\
   \hline
  \endhead 
   \hline
  \endfoot 
    $  1 $& $         .....  $ & $         .....  $ & $          25.79 $ & $          24.52 $ & $          23.66 $ & $          23.83 $ & $          22.84 $ & $          23.31 $ & $          23.12 $ & $         .....  $ & $          22.79 $ & $          22.84 $ & $          22.83 $ & $         <21.36 $ & $         .....   $ \\
 $    $& $                $ & $                $ & $       \pm 0.32 $ & $       \pm 0.16 $ & $       \pm 0.22 $ & $       \pm 0.07 $ & $       \pm 0.08 $ & $       \pm 0.12 $ & $       \pm 0.06 $ & $                $ & $       \pm 0.14 $ & $       \pm 0.13 $ & $       \pm 0.16 $ & $                $ & $                 $ \\
 $  2 $& $         .....  $ & $         .....  $ & $          25.15 $ & $          23.21 $ & $          22.94 $ & $          22.33 $ & $          21.90 $ & $          22.04 $ & $          21.76 $ & $          21.96 $ & $         .....  $ & $          21.34 $ & $          21.20 $ & $          21.40 $ & $         <21.01  $ \\
 $    $& $                $ & $                $ & $       \pm 0.17 $ & $       \pm 0.09 $ & $       \pm 0.09 $ & $       \pm 0.02 $ & $       \pm 0.03 $ & $       \pm 0.03 $ & $       \pm 0.02 $ & $       \pm 0.13 $ & $                $ & $       \pm 0.04 $ & $       \pm 0.05 $ & $       \pm 0.22 $ & $                 $ \\
 $  3 $& $         .....  $ & $         .....  $ & $         .....  $ & $         .....  $ & $         .....  $ & $         <24.18 $ & $          25.08 $ & $         .....  $ & $          24.33 $ & $         .....  $ & $         .....  $ & $          23.63 $ & $         <22.27 $ & $         <21.22 $ & $         .....   $ \\
 $    $& $                $ & $                $ & $                $ & $                $ & $                $ & $                $ & $       \pm 0.19 $ & $                $ & $       \pm 0.15 $ & $                $ & $                $ & $       \pm 0.21 $ & $                $ & $                $ & $                 $ \\
\end{longtable}
\end{landscape}
 \twocolumn

\label{lastpage}

\end{document}